\documentclass[11pt]{article}

\usepackage[margin=1in]{geometry}
\usepackage{amsmath,amssymb,amsfonts,bm}
\usepackage{graphicx}
\usepackage{booktabs}
\usepackage{authblk}
\usepackage{hyperref}
\usepackage{cleveref}
\usepackage{color}
\usepackage{placeins}
\usepackage{epstopdf}
\usepackage{subfigure}
\title{AQP4 Regulates Gap-Dominated Glymphatic Clearance through Dynamic Gliovascular Coupling}

\author[1]{Tanran Zhang}
\affil[1]{School of Mathematical Sciences, Soochow University,  Jiangsu, China}
\author[2]{Huaxiong Huang}
\affil[2]{Zu Chongzhi Center, Duke Kunshan University, Kunshan, Jiangsu, China}
\author[3]{Robert Eisenberg }
\affil[3]{Department of Physiology and Biophysics, Rush University, Chicago 60612 USA}
\author[2]{Shixin Xu\thanks{Corresponding author: \href{mailto:shixin.xu@dukekunshan.edu.cn}{shixin.xu@dukekunshan.edu.cn}}}

\date{}

\begin{document}

\maketitle

\begin{abstract}
    Aquaporin-4 (AQP4) is enriched at perivascular astrocytic endfeet, and impaired AQP4 function or localization is associated with reduced glymphatic transport. However, mechanical models suggest that pressure-driven exchange across the gliovascular interface occurs predominantly through inter-endfoot gaps rather than directly through the AQP4-rich membrane. How AQP4 regulates clearance in such a gap-dominated system remains unclear. We develop a reduced arterial PVS-ECS-venous PVS model coupling vascular deformation, AQP4-mediated endfoot water exchange, dynamic inter-endfoot gap regulation, and tracer transport. Cardiac-like oscillations generate strong bidirectional exchange but weak net clearance, whereas asymmetric vasodilation enhances directional transport by suppressing recovery-phase backflow. Dynamic gap regulation provides additional hydraulic rectification. Although gap-mediated flux is much larger than direct AQP4-mediated flux, PVS-facing AQP4 substantially affects clearance by altering the pressure--volume balance and hence the driving force for the dominant gap pathway. Reducing PVS-facing AQP4 permeability, or redistributing AQP4 away from the PVS-facing membrane at fixed total conductance, reduces cumulative venous output by about 40\%. Aging-like changes in vascular motion, PVS mechanical coupling, and perivascular AQP4 enrichment further compound this impairment. These results suggest that AQP4 regulates gap-dominated glymphatic clearance through dynamic hydraulic coupling at the gliovascular interface.
\end{abstract}

\section{Introduction}

 The glymphatic system has been proposed as a brain-wide transport
pathway in which cerebrospinal fluid (CSF) enters along periarterial
perivascular spaces (PVSs), exchanges with interstitial fluid in the
extracellular space (ECS), and contributes to the clearance of
metabolic waste
\cite{Iliff2012,Bohr2022,Jessen2015,mestre2018flow}.
Impaired glymphatic transport has been associated with reduced
clearance of amyloid-\(\beta\), tau, and other neurotoxic species, and
has been implicated in aging, Alzheimer's disease, traumatic brain
injury, stroke, and sleep-related disturbances of waste removal
\cite{Iliff2014,Silva2021,mestre2018flow,khan2026impaired}.

Aquaporin-4 (AQP4), which is highly enriched in astrocytic endfeet
surrounding cerebral vessels, is a central component of the glymphatic
hypothesis \cite{mestre2018flow,Silva2021}. Experimental evidence
suggests that both AQP4 abundance and its perivascular polarization
contribute to efficient cerebrospinal fluid-interstitial fluid
(CSF-ISF) exchange. In the original glymphatic study, AQP4-deficient
mice exhibited markedly reduced CSF tracer influx and an approximately
\(55\%\) reduction in amyloid-\(\beta\) clearance compared with
wild-type controls \cite{Iliff2012}. Subsequent studies have reported
impaired tracer transport or solute clearance following AQP4 deletion,
pharmacological inhibition, or loss of perivascular polarization in
aging and neurological disease models
\cite{Iliff2014,Simon2022,kress2014impairment,
bojarskaite2024role}. Although the reported magnitude of the effect
varies across experimental protocols, the available evidence supports
an important regulatory role for AQP4 in glymphatic transport.

The physical mechanism underlying this regulation, however, remains
uncertain. One hypothesis is that AQP4 provides the
dominant pathway for water to cross the astrocytic endfoot layer from
the PVS to the ECS. This interpretation is appealing because AQP4 is
highly enriched on perivascular endfoot membranes and forms
highly water-permeable channels. Loss of AQP4 expression or
polarization would then directly reduce trans-endfoot hydraulic
permeability and impede PVS-ECS exchange.

An alternative possibility is that most pressure-driven water transport
crosses the gliovascular interface through extracellular gaps between
neighboring endfeet, while AQP4 regulates this larger pathway
indirectly by modifying endfoot water balance, compartment pressures,
and the pressure difference across the gaps \cite{Causemann2026}. The central mechanistic
question is therefore not simply whether AQP4 affects glymphatic
clearance, but how it does so: does AQP4 carry the dominant
trans-endfoot water flux, or does it regulate the hydraulic conditions
governing a larger gap-mediated pathway?

Mathematical modeling provides a means of distinguishing these
mechanisms. At the local PVS scale, spatially resolved models have
examined how arterial-wall motion, PVS geometry, and tissue deformation
generate fluid and solute transport. Asgari et al.\ showed that
oscillatory flow in a deforming periarterial space can enhance solute
dispersion without requiring substantial unidirectional bulk flow
\cite{Asgari2016}. Using anatomically informed deforming geometries,
Kedarasetti et al.\ found that physiological cardiac pulsations
generate predominantly oscillatory CSF motion and are insufficient by
themselves to account for the magnitude of observed directional flow
\cite{Kedarasetti2020}. Roman\`o et al.\ demonstrated that
perivascular exchange depends strongly on tissue elasticity,
permeability, and the phase relation between arterial and outer-boundary
motion \cite{Romano2020}. Together, these studies show that vascular
motion need not translate directly into net transport; instead, the
outcome depends on waveform timing, boundary mechanics, and hydraulic
coupling \cite{Bohr2022,Kedarasetti2022}.

More recent high-fidelity simulations that resolve individual
astrocytic endfeet suggest that pressure-driven exchange across the
gliovascular interface may occur predominantly through extracellular
inter-endfoot gaps rather than directly through the AQP4-rich endfoot
membrane \cite{Causemann2026}. This finding creates an apparent
mechanistic tension with experimental observations showing substantial
reductions in tracer influx and clearance after AQP4 disruption. If the
direct hydrostatic flux through the AQP4-rich membrane is comparatively
small, AQP4 must influence clearance either through its effects on
local pressure and volume regulation or through its interaction with
the dominant gap-mediated pathway.

Spatially resolved models provide detailed information about local
fluid mechanics, but their restricted domains and computational cost
make it difficult to follow long-time tracer redistribution through the
full periarterial-parenchymal-perivenous pathway. Reduced-order and
compartmental models provide a complementary level of description. By
replacing spatial fields with a small number of pressures, volumes,
concentrations, and hydraulic resistances, they permit long-time
simulations and systematic parameter studies
\cite{DavoodiBojd2019,Tithof2022}. Most existing reduced models,
however, prescribe the PVS-parenchyma conductance, focus on steady
hydraulic resistance, or infer effective transport rates from imaging
data. Few explicitly couple vascular waveform, geometry-induced PVS
deformation, AQP4-mediated endfoot water exchange, dynamic
inter-endfoot gap conductance, and downstream tracer clearance within
a single time-dependent framework.

Three related questions therefore remain unresolved. First, how can a
comparatively small AQP4-mediated water flux substantially influence
clearance when instantaneous flow across the gliovascular interface is
dominated by the gap pathway? Second, how does vascular waveform shape
interact with AQP4-dependent endfoot dynamics and gap regulation to
convert bidirectional fluid motion into net PVS-ECS transport and
downstream clearance? Cardiac pulsation may produce large forward and
backward displacements with strong cancellation, whereas slow or
temporally asymmetric vasodilation may generate more directional
transport by reducing recovery-phase backflow. Third, how do
aging-associated changes in vascular motion, PVS mechanical coupling,
and perivascular AQP4 polarization combine to impair this transport
system? Aging simultaneously reduces arterial-wall deformation,
remodels the periarterial environment, and disrupts perivascular AQP4
localization
\cite{kress2014impairment,mestre2018flow,
mestre2022periarteriolar,khan2026impaired}, but the combined effects of
these changes on dynamic gap regulation and venous-directed clearance
remain poorly understood.

To address these questions, we develop a reduced
arterial-ECS-chain-venous multicompartment model that explicitly
distinguishes AQP4-mediated trans-endfoot water exchange from
inter-endfoot gap flow while retaining their coupling through
compartment pressures and endfoot volume regulation. The model combines
vascular forcing, geometry-induced PVS deformation, dynamic gap
opening, AQP4-mediated water exchange at the endfoot membranes, and
passive tracer transport. It also allows total functional
AQP4-associated conductance and perivascular AQP4 polarization to be
varied separately. This model therefore complements high-fidelity gliovascular simulations \cite{Causemann2026}
by showing how a gap-dominated local exchange mechanism can be linked
to dynamic rectification, AQP4 polarization, and cumulative downstream
clearance at the network scale.

The arterial-side network contains an arterial PVS, an astrocytic
endfoot compartment, and an arterial ECS compartment. These are
connected through intermediate and venous-side ECS compartments to a
venous PVS and a downstream outlet, as illustrated in
Fig.~\ref{fig:schematic}. This structure distinguishes local
bidirectional PVS-ECS exchange from net gap-mediated transport, tracer
redistribution through the ECS chain, venous PVS accumulation, and
final downstream output. We use the model to determine whether AQP4 can
regulate clearance even when it does not carry the dominant
instantaneous water flux, to identify how vascular waveform and
gap-response timing produce hydraulic rectification, and to quantify
the combined effects of impaired vascular forcing, altered PVS
mechanics, and loss of AQP4 polarization in representative aging-like
conditions.

The remainder of the paper is organized as follows.
Section~\ref{sec:model} introduces the multicompartment geometry,
pressure-volume relations, water-flux laws, tracer transport equations,
dynamic-gap formulation, and vascular forcing protocols.
Section~\ref{sec:results} presents waveform-dependent exchange and
clearance, the dynamic-gap mechanism and response-time sensitivity,
the effects of AQP4 abundance and polarization, and representative
aging-like simulations. Section~\ref{sec:discussion} discusses the
mechanistic implications, limitations, experimentally testable
predictions, and relevance of the results to aging-associated
glymphatic impairment.

\section{Arterial-ECS-chain-venous clearance model}\label{sec:model}

\begin{figure}[!htp]
    \centering
    \includegraphics[width=0.65\linewidth]{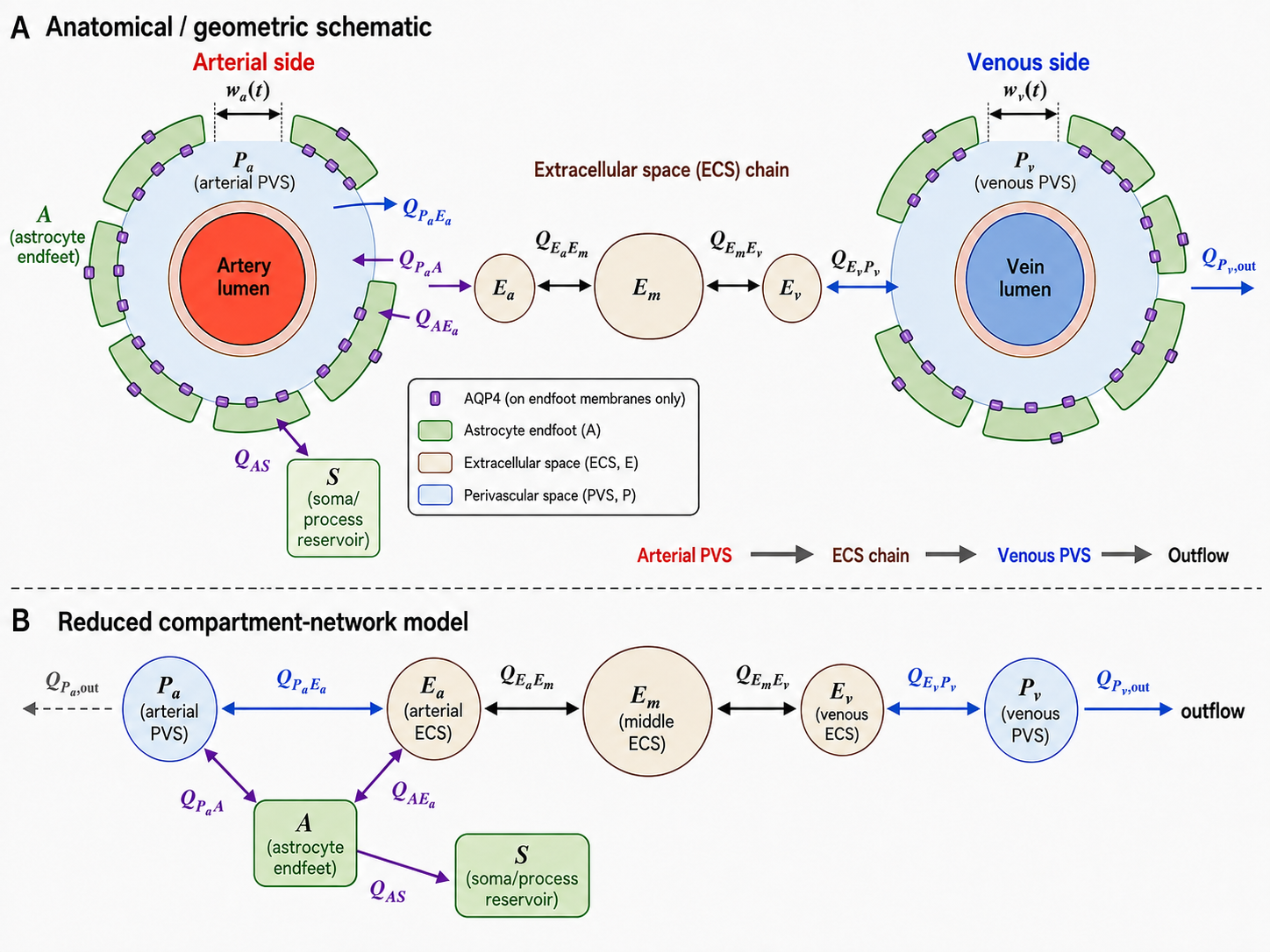}
    \caption{Schematic illustration of the reduced arterial-ECS-chain-venous model. (A) Multi-compartment geometric representation of the arterial PVS, astrocytic endfoot layer, extracellular-space compartments, and venous PVS. AQP4 channels are located on the endfoot membranes, while inter-endfoot gaps provide a direct gap-mediated pathway between the PVS and ECS. (B) Reduced compartment-network representation showing the water and tracer exchange pathways among $P_a$, $A$, $E_a$, $E_m$, $E_v$, and $P_v$, together with downstream venous outflow. The glymphatic system description of the anatomy can be found in \cite{Jessen2015}. }
    \label{fig:schematic}
\end{figure}

As shown in Fig.\ref{fig:schematic}, the model consists of six fluid compartments: the arterial perivascular space \(P_a\), the arterial-side astrocytic endfoot \(A\), three extracellular-space compartments \(E_a,E_m,E_v\), and the venous perivascular space \(P_v\). 
Here \(E_a\) denotes the arterial-side ECS, \(E_m\) represents a middle ECS buffering region, and \(E_v\) denotes the venous-side ECS. 
The middle ECS compartment is assigned a larger reference volume than the two boundary ECS compartments, reflecting its role as a parenchymal buffering region. 
The model network is shown in Fig. \ref{fig:schematic}(b).
This reduced network allows us to separate arterial CSF influx, interstitial buffering, venous-side PVS storage, and true downstream venous clearance.



The effective compartment volumes $V_i$ each have two components. One component is a ``baseline" reference volume  \(V_i^{\mathrm{ref}}\), independent of pressure \(p_i(t)\). The other component of each volume depends on pressure as indicated. 
\[
V_{P_a}(t)
=
V_{P_a}^{\mathrm{geom}}(t)
+
C_{P_a}\left(p_{P_a}-p_{P_a,0}\right),
\]
\[
V_A(t)
=
V_A^{\mathrm{ref}}
+
C_A\left(p_A-p_{A,0}\right),
\]
\[
V_{E_a}(t)
=
V_{E_a}^{\mathrm{ref}}
+
C_{E_a}\left(p_{E_a}-p_{E_a,0}\right),
\]
\[
V_{E_m}(t)
=
V_{E_m}^{\mathrm{ref}}
+
C_{E_m}\left(p_{E_m}-p_{E_m,0}\right),
\]
\[
V_{E_v}(t)
=
V_{E_v}^{\mathrm{ref}}
+
C_{E_v}\left(p_{E_v}-p_{E_v,0}\right),
\]
and
\[
V_{P_v}(t)
=
V_{P_v}^{\mathrm{ref}}
+
C_{P_v}\left(p_{P_v}-p_{P_v,0}\right).
\]
 where  \(p_{i,0}\)
is the corresponding pressure in the reference equilibrium state.
  The effective compliance is related to the compartment stiffness
by
\[
C_i
=
\frac{V_i^{\mathrm{ref}}}{K_i}.
\]
The compliance term represents the change in stored
fluid volume caused by a pressure deviation from the reference state.
For the arterial PVS, \(V_{P_a}^{\mathrm{geom}}(t)\) additionally
accounts for the geometric deformation induced by the prescribed
vascular motion.

In the simulations, the middle ECS compartment is chosen to be larger
than the arterial-side and venous-side ECS compartments because it
represents the broader parenchymal interstitial space connecting the
localized periarterial and perivenous exchange regions. It should be
interpreted as a lumped storage and transport reservoir rather than as
a sharply bounded anatomical compartment \cite{Jessen2015}.

The geometric arterial PVS volume is
\[
V_{P_a}^{\mathrm{geom}}(t)
=
\pi L_a\left(R_o(t)^2-R_v(t)^2\right),
\]
where the outer PVS radius is modeled as
\[
R_o(t)=R_{o0}+\eta_o\left(R_v(t)-R_{v0}\right),
\]
and the arterial vessel radius is prescribed by
\[
R_v(t)=R_{v0}\left(1+\epsilon f(t)\right).
\]
Here \(\eta_o\) measures how strongly the outer PVS boundary follows the vessel wall motion and 
 \(f(t)\) represents the vascular waveform and \(\epsilon\) is the relative amplitude.



In this work, the prescribed arteriolar waveforms are motivated by three
complementary classes of in vivo experiments: cardiac pulsation,
spontaneous low-frequency vasomotion, and transient functional
vasodilation. These vascular signals differ not only in their
timescales, but also in the temporal relationship between dilation and
recovery. The purpose of comparing them is   to determine how
waveform frequency and temporal asymmetry affect PVS compression,
forward-backward fluid cancellation, dynamic gap regulation, and
downstream solute clearance.

The waveforms introduced below are idealized representations of the
corresponding experimental observations, rather than direct fits to a
particular vessel trace. This abstraction allows the effects of
waveform shape to be separated from differences in vessel size,
forcing amplitude, experimental preparation, and measurement protocol.
In particular, the symmetric and asymmetric slow-wave protocols are
assigned the same period, maximum dilation, and integrated waveform
magnitude, so that their comparison isolates the effect of temporal
asymmetry.  

\paragraph{Cardiac-like zero-mean waveform.} Two-photon particle-tracking experiments have shown that periarterial
CSF motion is synchronized with cardiac-driven arterial-wall motion
\cite{mestre2018flow}. We include a rapid periodic waveform to determine
whether cardiac pulsation produces net directional transport or
primarily large forward-backward exchange with substantial
cancellation.
 
\begin{equation}
    f_c(t)=\sin\left(\frac{2\pi t}{T_c}\right).
    \label{eq:cardiac}
\end{equation}
where we set the period $T_c =0.2 s$   corresponding to a frequency of
\(5\,\mathrm{Hz}\) or \(300\) beats/min. This value represents a
rodent-like cardiac frequency. 
This waveform has positive and negative phases, representing dilation and contraction around the baseline radius.

\paragraph{Symmetric slow vasomotion.} 
Slow arteriolar diameter oscillations near \(0.1\,\mathrm{Hz}\) have
been observed in awake mice and associated with enhanced paravascular
solute clearance \cite{van2020vasomotion}. The symmetric waveform represents
an idealized dilation-recovery cycle and tests whether dynamic
pressure-conductance coupling can rectify an otherwise temporally
balanced vascular signal.
\begin{equation}
    \tilde{f}_s(\phi)
    =  \exp\left[-\frac{(\phi-\phi_0)^2}{2\sigma^2}\right], ~f_s(\phi)= 
    \frac{\tilde{f}_s(\phi)}{
    \max_{\phi\in[0,1]}
     \left(\tilde{f}_s(\phi)\right)},
    \label{eq:sym_wave}
\end{equation}
where 
\(
\phi=\frac{t \bmod T_p}{T_p}
\)
be the phase within one period \(T_p\),
The denominator normalizes the waveform so that
\[
\max_{\phi\in[0,1]} f_s(\phi)=1.
\]
For the parameter choices used here, \(\phi_0\in[0,1]\), so the maximum occurs at \(\phi=\phi_0\) and the denominator equals one.

\begin{figure}
    \centering
    \includegraphics[width=0.75\linewidth]{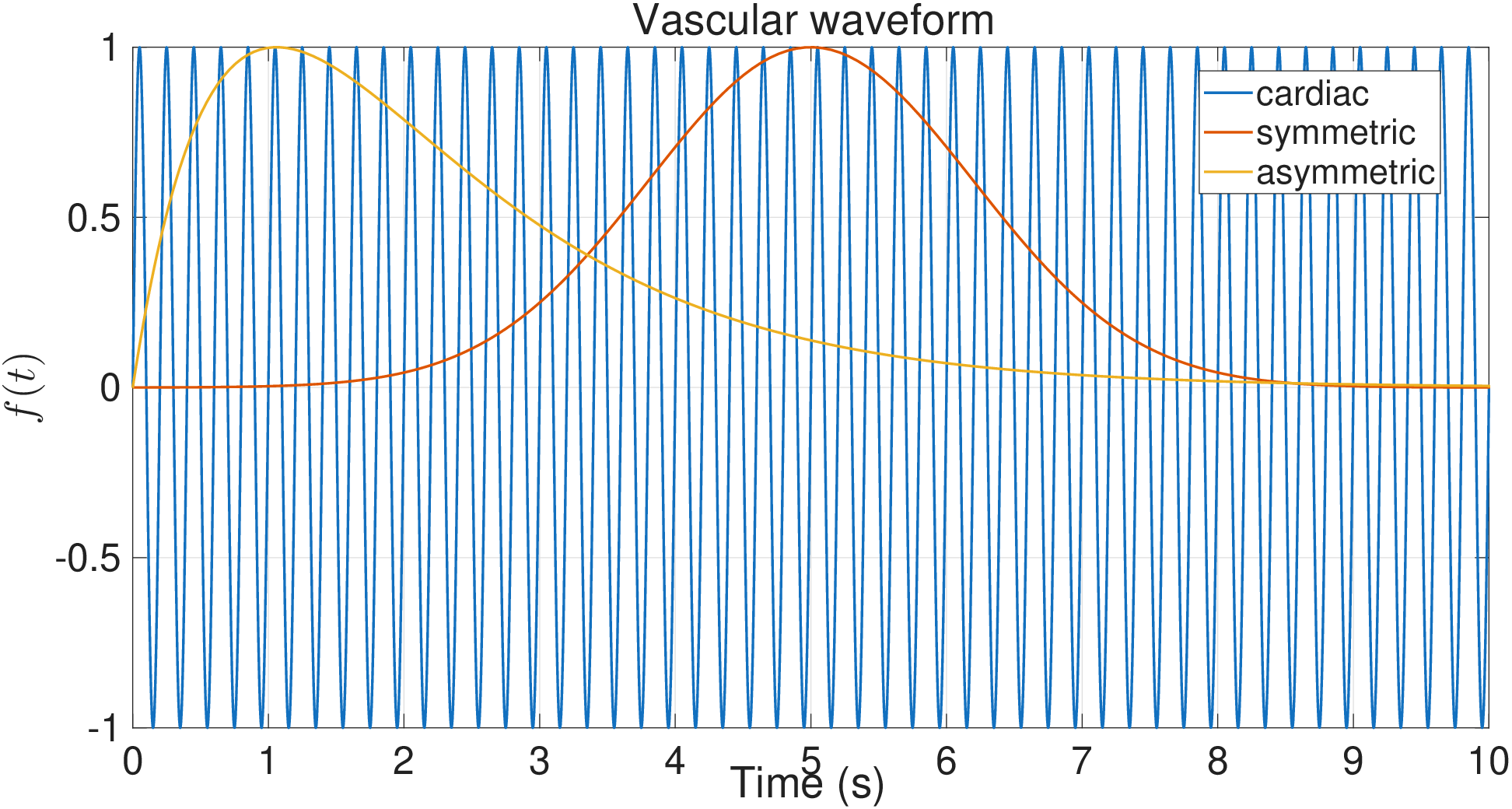}
    \caption{Illustration of the three vascular waveforms. }
    \label{fig:waveform}
\end{figure}

\paragraph{Asymmetric vasodilation.}
Sensory-stimulation experiments have shown that functional arteriolar
dilation can enhance periarterial CSF influx and solute clearance
\cite{HolsteinRonsbo2023}. We represent this response by a waveform with
unequal dilation and recovery phases to test whether temporal asymmetry
reduces reverse flow and promotes net downstream clearance.
\begin{equation}
   \tilde{f}_a(\phi)
= e^{-\phi/\tau_d}-e^{-\phi/\tau_r},~ f_a(\phi) =
    \frac{ \tilde{f}_a(\phi)
    }{
    \max_{\phi\in[0,1]}
    \left( \tilde{f}_a(\phi)\right)
    },
    \qquad
    \tau_r<\tau_d.
    \label{eq:asym_wave}
\end{equation}
Here \(\tau_r\) controls the rise time and \(\tau_d\) controls the decay time. 
The symmetric and asymmetric slow waveforms are assigned the same
period, peak dilation, and integrated waveform magnitude as shown in Fig. \ref{fig:waveform}. Their
comparison therefore isolates the effect of temporal asymmetry rather
than differences in overall forcing strength. 

\subsection{Pressure dynamics and water fluxes}

If all water fluxes
are signed, with \(Q_{ij}>0\) denoting flow from compartment \(i\) to
compartment \(j\), then based on the conservation law, the arterial PVS pressure satisfies
\begin{equation}
C_{P_a}\frac{dp_{P_a}}{dt}
=
S_a
-
Q_{P_aE_a}
-
Q_{P_aA}
-
Q_{P_a,\mathrm{out}},
\label{eq:Pa_pressure}
\end{equation}
where
\begin{equation}\label{eq:S_a}
S_a
=
-\frac{dV_{P_a}^{\mathrm{geom}}}{dt}
\end{equation}
is the prescribed geometric compression source. Thus, \(S_a>0\)
corresponds to compression of the arterial PVS and produces a positive
pressure-driving contribution.

The remaining compartment pressures satisfy
\begin{align}
C_A\frac{dp_A}{dt}
&=
Q_{P_aA}
-
Q_{AE_a}
-
Q_{AS},
\\
C_{E_a}\frac{dp_{E_a}}{dt}
&=
Q_{P_aE_a}
+
Q_{AE_a}
-
Q_{E_aE_m},
\\
C_{E_m}\frac{dp_{E_m}}{dt}
&=
Q_{E_aE_m}
-
Q_{E_mE_v},
\\
C_{E_v}\frac{dp_{E_v}}{dt}
&=
Q_{E_mE_v}
-
Q_{E_vP_v},
\\
C_{P_v}\frac{dp_{P_v}}{dt}
&=
Q_{E_vP_v}
-
Q_{P_v,\mathrm{out}}.
\label{eq:pressure_system}
\end{align}


For the direct arterial PVS-ECS pathway, under the parallel-plate Poiseuille
approximation, the corresponding hydraulic conductance satisfies  
\[
G_{P_aE_a}
=
G_{a,\mathrm{gap},0}w_a^3
\]
because the conductance of a narrow slit scales as the third power of
its width \cite{koch2023estimates}.

Water exchange across the PVS-facing and ECS-facing endfoot membranes
would, in general, depend on both hydrostatic and osmotic pressure
differences:
\[
Q_{P_aA}
=
L_{P_aA}^{\mathrm{eff}}
\left[
(p_{P_a}-p_A)-\Delta\Pi_{P_aA}\right],~
Q_{AE_a}
=
L_{AE_a}^{\mathrm{eff}}
\left[
(p_A-p_{E_a})-\Delta\Pi_{AE_a}
\right].
\]

The present study focuses on transport generated by vascular deformation
and hydrostatic pressure coupling.  The possible role of spatially nonuniform osmotic forces (such as those that are so important in the kidney) in glymphatic clearance needs further separate investigation.  We
therefore assume that baseline osmotic contributions are balanced in the
reference state and neglect their dynamic variations.
\[
\Delta\Pi_{P_aA}
=
\Delta\Pi_{AE_a}
=
0.
\]
This assumption isolates the hydrostatic effects of vascular waveform,
PVS deformation, and hydraulic network coupling; it does not imply that
osmotic forces are physiologically negligible.

The effective membrane conductances are defined as follows 
\[
L_{P_aA}^{\mathrm{eff}}
=
L_{P_aA}^{\mathrm{bg}}
+
L_{\mathrm{AQP4}}^{PVS},
\]
and
\[
L_{AE_a}^{\mathrm{eff}}
=
L_{AE_a}^{\mathrm{bg}}
+
L_{\mathrm{AQP4}}^{ECS}.
\]
Here, \(L_{P_aA}^{\mathrm{bg}}\) and
\(L_{AE_a}^{\mathrm{bg}}\) represent non-AQP4 background water
conductance. \(L_{\mathrm{AQP4}}^{PVS}\) and \(L_{\mathrm{AQP4}}^{ECS} \)   are the water conductance induced by AQP4. 
We introduce
\(\rho_{\mathrm{AQP4}}\in[0,1]\), defined as the fraction of the total
AQP4-associated hydraulic conductance assigned to the PVS-facing
endfoot membrane. The remaining fraction is assigned to the
ECS-facing membrane:
\[
L_{\mathrm{AQP4}}^{PVS}
=
\rho_{\mathrm{AQP4}}
L_{\mathrm{AQP4}}^{\mathrm{tot}},
\qquad
L_{\mathrm{AQP4}}^{ECS}
=
\left(1-\rho_{\mathrm{AQP4}}\right)
L_{\mathrm{AQP4}}^{\mathrm{tot}}.
\]
Thus,
\[
L_{\mathrm{AQP4}}^{PVS}
+
L_{\mathrm{AQP4}}^{ECS}
=
L_{\mathrm{AQP4}}^{\mathrm{tot}},
\]
so that the total functional AQP4-associated conductance is conserved
throughout the redistribution study.

 We set
\[
L_{\mathrm{AQP4}}^{\mathrm{tot}}=0.05,
\qquad
L_{P_aA}^{\mathrm{bg}}
=
L_{AE_a}^{\mathrm{bg}}
=
0.005.
\]


The remaining water fluxes are  assumed to be driven primarily by
hydrostatic pressure differences and is described by a Darcy-type
relation,
\[
Q_{AS}
=
G_{AS}(p_A-p_S),
\]
\[
Q_{E_aE_m}
=
G_{E_aE_m}(p_{E_a}-p_{E_m}),
\qquad
Q_{E_mE_v}
=
G_{E_mE_v}(p_{E_m}-p_{E_v}),
\]
\[
Q_{E_vP_v}
=
G_{E_vP_v}(p_{E_v}-p_{P_v}),
\]
and
\[
Q_{P_v,\mathrm{out}}
=
G_{P_v,\mathrm{out}}
(p_{P_v}-p_{P_v,\mathrm{out}}).
\]



 


\subsection{Dynamic arterial gap regulation}

The astrocytic endfoot sheath forms a nearly continuous boundary between
the arterial PVS and the surrounding ECS, with extracellular transport
occurring through narrow clefts between neighboring endfeet \cite{Iliff2014,mathiisen2010perivascular,koch2023estimates} as shown in Fig. \ref{fig:schematic}
. Because the hydraulic conductance of
these clefts is highly sensitive to their width, small deformations of
the endfoot sheath may substantially alter PVS-ECS water exchange
\cite{koch2023estimates}. We therefore introduce a dynamic,
dimensionless arterial gap factor \(w_a(t)\), governed by the relaxation
law
\begin{equation}
\tau_w\frac{dw_a}{dt}
=
w_{a,*}-w_a,
\end{equation}
where   \(w_{a,*}\) is
the instantaneous target gap factor. The relaxation time \(\tau_w\) is an effective
response time of the inter-endfoot gap. It collectively represents the
viscoelastic response of the astrocytic endfoot sheath, endfoot-volume
regulation, and local hydraulic equilibration within the PVS-ECS
interface. These processes may be influenced by cytoskeletal mechanics,
membrane water permeability, ion-dependent volume regulation, and the
hydraulic properties of the surrounding extracellular matrix. Because
direct in vivo measurements of inter-endfoot gap response times are not
currently available, \(\tau_w\) is treated as an effective parameter
and examined through sensitivity analysis.

To represent the combined effects of vascularly induced deformation and
endfoot-volume changes, we set
\begin{equation}
w_{a,*}
=
w_0
+
g_{\mathrm{open}}
\frac{S_a^+}{S_{a,0}}
-
g_{\mathrm{volA}}s_A,
\end{equation}
where
\[
S_a^+=\max(S_a,0),
\qquad
s_A=\frac{V_A-V_A^{\mathrm{ref}}}{V_A^{\mathrm{ref}}}.
\]
Here, \(w_0\) is the resting gap factor, \(S_a\) is the mechanical PVS
compression source defined in Eq. \eqref{eq:S_a}, and \(S_{a,0}\) is a normalization scale, taken as
the maximum value of \(\lvert S_a\rvert\) over one waveform period.

The first feedback term represents mechanically induced enhancement of
the effective inter-endfoot gap conductance during arterial PVS
compression. Arterial dilation displaces the inner PVS boundary,
compresses the PVS, and mechanically deforms the surrounding astrocytic
endfoot sheath. Recent high-fidelity poroelastic simulations predict
that this deformation includes tangential stretching of the endfoot
sheath and that the resulting pressure-driven exchange occurs
predominantly through inter-endfoot gaps
\cite{Causemann2026}.Motivated by these observations, we use the positive part of the
compression source, \(S_a^+\), as a phenomenological approximation of
the component of vascular deformation that promotes effective opening,
and therefore increased hydraulic conductance, of the inter-endfoot gap
pathway.  The positive-part operator restricts
this active opening response to the compression phase; during the
recovery phase, the gap relaxes toward its resting state.

The second feedback term accounts for the effect of endfoot-volume
changes. Astrocytic endfeet are highly water permeable and can undergo
significant volume changes, with AQP4 contributing to astrocytic water
transport and volume regulation
\cite{thrane2011critical,rosic2019aquaporin,sucha2022absence}. Geometrically, expansion of the
neighboring endfeet reduces the extracellular space available between
them and is therefore expected to decrease the effective width and
hydraulic conductance of the inter-endfoot cleft. Accordingly,
\(s_A>0\) produces a reduction in \(w_{a,*}\), whereas endfoot shrinkage,
\(s_A<0\), tends to increase the effective gap factor.

Direct in vivo measurements of time-dependent inter-endfoot gap width
are currently unavailable. The above relation should therefore be
interpreted as a physiologically motivated reduced constitutive law
rather than a quantitatively established microscopic law. The
coefficients \(g_{\mathrm{open}}\), \(g_{\mathrm{volA}}\), and
\(\tau_w\) characterize the effective mechanical sensitivity,
volume-feedback strength, and response time, respectively, and their
influence is examined through sensitivity analysis.

\subsection{Solute transport}
We consider the transport of a passive solute through the
multicompartment system. In general, this variable may represent a
dilute extracellular nonphysiological substance transported by the interstitial fluid.
Because the present study focuses on glymphatic clearance, we interpret
the modeled solute as an experimentally detectable extracellular tracer
and refer to it as the tracer hereafter.

The tracer is transported by advection associated with
intercompartmental water fluxes and by effective diffusive exchange
between neighboring ECS compartments. For the relatively large
extracellular tracer considered here, direct passage through the
astrocytic endfoot membrane is neglected. AQP4 therefore affects tracer
transport only indirectly through its effects on water exchange, compartment
pressures and volumes, and dynamic gap conductance.

Let \(M_l\) denote the total tracer mass in the \(l\)-th compartment.
The corresponding tracer concentration is defined by  
 
\[
c_{l}=\frac{M_{l}}{V_{l}},\qquad
l=P_a, E_a, E_m, E_v,P_v
\]

For any water flux \(Q_{ij}\), the advective tracer flux from \(i\) to \(j\) is discretized using an upwind form:
\[
J_{ij}^{\mathrm{adv}}
=
Q_{ij}^{+}c_i
-
Q_{ij}^{-}c_j,
\]
where
\[
Q_{ij}^{+}=\max(Q_{ij},0),
\qquad
Q_{ij}^{-}=\max(-Q_{ij},0).
\]
Therefore \(J_{ij}>0\) denotes net tracer flux from \(i\) to \(j\).

The tracer flux from arterial PVS to arterial-side ECS is
\[
J_{P_aE_a}
=
Q_{P_aE_a}^{+}c_{P_a}
-
Q_{P_aE_a}^{-}c_{E_a}
+
D_{P_aE_a}\left(c_{P_a}-c_{E_a}\right).
\]
The ECS-chain tracer fluxes are
\[
J_{E_aE_m}
=
Q_{E_aE_m}^{+}c_{E_a}
-
Q_{E_aE_m}^{-}c_{E_m}
+
D_{E_aE_m}\left(c_{E_a}-c_{E_m}\right),
\]
\[
J_{E_mE_v}
=
Q_{E_mE_v}^{+}c_{E_m}
-
Q_{E_mE_v}^{-}c_{E_v}
+
D_{E_mE_v}\left(c_{E_m}-c_{E_v}\right).
\]
The tracer flux from venous-side ECS to venous PVS is
\[
J_{E_vP_v}
=
Q_{E_vP_v}^{+}c_{E_v}
-
Q_{E_vP_v}^{-}c_{P_v}
+
D_{E_vP_v}\left(c_{E_v}-c_{P_v}\right).
\]

The arterial and venous boundary tracer fluxes are
\[
J_{P_a,\mathrm{out}}
=
Q_{P_a,\mathrm{out}}^{+}c_{P_a}
-
Q_{P_a,\mathrm{out}}^{-}c_{P_a,\mathrm{out}},
\]
and
\[
J_{P_v,\mathrm{out}}
=
Q_{P_v,\mathrm{out}}^{+}c_{P_v}
-
Q_{P_v,\mathrm{out}}^{-}c_{P_v,\mathrm{out}}.
\]

The tracer mass balances are
\begin{eqnarray}
    \frac{dM_{P_a}}{dt}&=&-J_{P_aE_a}-J_{P_a,\mathrm{out}},\\
\frac{dM_{E_a}}{dt}&=&J_{P_aE_a}-J_{E_aE_m},\\
\frac{dM_{E_m}}{dt}&=&J_{E_aE_m}-J_{E_mE_v},\\
\frac{dM_{E_v}}{dt}&=&J_{E_mE_v}-J_{E_vP_v},\\
\frac{dM_{P_v}}{dt}&=&J_{E_vP_v}-J_{P_v,\mathrm{out}}.
\end{eqnarray}

In the present study, local tracer degradation within the ECS
compartments is neglected; Thus, tracer mass can therefore decrease only through intercompartmental
transport and boundary outflow.
 




\begin{table}[htbp]
\centering
\caption{Geometric, storage, and hydraulic parameters in the extended arterial-ECS-chain-venous model.}
\label{tab:model_parameters}
\begin{tabular}{llll}
\hline
Parameter & Description & Baseline value & Unit \\
\hline
\(R_{v0}\) & Baseline arterial vessel radius & \(10\) & \(\mu\mathrm{m}\) \\
\(R_{o0}\) & Baseline outer arterial PVS radius & \(12\) & \(\mu\mathrm{m}\) \\
\(L_a\) & Arterial segment length & \(100\) & \(\mu\mathrm{m}\) \\
\(\epsilon\) & Relative radius-change amplitude & \(0.02\) & - \\
\(\eta_o\) & Outer-wall motion factor & \(0.2\) & - \\
\(V_{P_a}^{\mathrm{ref}}\) & Reference arterial PVS volume & \(\pi L_a(R_{o0}^2-R_{v0}^2)\) & \(\mu\mathrm{m}^3\) \\
\(V_A^{\mathrm{ref}}\) & Reference endfoot volume & \(0.6V_{P_a}^{\mathrm{ref}}\) & \(\mu\mathrm{m}^3\) \\
\(V_{E_a}^{\mathrm{ref}}\) & Reference arterial-side ECS volume & \(2V_{P_a}^{\mathrm{ref}}\) & \(\mu\mathrm{m}^3\) \\
\(V_{E_m}^{\mathrm{ref}}\) & Reference middle ECS volume & \(4V_{P_a}^{\mathrm{ref}}\) & \(\mu\mathrm{m}^3\) \\
\(V_{E_v}^{\mathrm{ref}}\) & Reference venous-side ECS volume & \(2V_{P_a}^{\mathrm{ref}}\) & \(\mu\mathrm{m}^3\) \\
\(V_{P_v}^{\mathrm{ref}}\) & Reference venous PVS volume & \(1.5V_{P_a}^{\mathrm{ref}}\) & \(\mu\mathrm{m}^3\) \\
\(K_{P_a}\) & Effective arterial PVS stiffness & \(1000\) & Pa \\
\(K_A\) & Effective endfoot stiffness & \(1000\) & Pa \\
\(K_{E_a}\) & Effective arterial-side ECS stiffness & \(500\) & Pa \\
\(K_{E_m}\) & Effective middle ECS stiffness & \(500\) & Pa \\
\(K_{E_v}\) & Effective venous-side ECS stiffness & \(500\) & Pa \\
\(K_{P_v}\) & Effective venous PVS stiffness & \(1000\) & Pa \\
\(G_{a,\mathrm{gap},0}\) & Baseline arterial gap conductance & \(1.0\) & \(\mu\mathrm{m}^3/(\mathrm{Pa}\cdot\mathrm{s})\) \\
\(L_{P_aA}^{\mathrm{bg}}
\) & PVS-endfoot backgroud water conductance & \(0.005\) & \(\mu\mathrm{m}^3/(\mathrm{Pa}\cdot\mathrm{s})\) \\
\(
L_{AE_a}^{\mathrm{bg}}\) & Endfoot-ECS backgroud  water conductance & \(0.005\) & \(\mu\mathrm{m}^3/(\mathrm{Pa}\cdot\mathrm{s})\) \\
\(G_{AS}\) & Endfoot-soma/process conductance & \(0.05\) & \(\mu\mathrm{m}^3/(\mathrm{Pa}\cdot\mathrm{s})\) \\
\(G_{E_aE_m}\) & \(E_a\)-\(E_m\) hydraulic conductance & \(10\) & \(\mu\mathrm{m}^3/(\mathrm{Pa}\cdot\mathrm{s})\) \\
\(G_{E_mE_v}\) & \(E_m\)-\(E_v\) hydraulic conductance & \(10\) & \(\mu\mathrm{m}^3/(\mathrm{Pa}\cdot\mathrm{s})\) \\
\(G_{E_vP_v}\) & \(E_v\)-venous PVS conductance & \(5\) & \(\mu\mathrm{m}^3/(\mathrm{Pa}\cdot\mathrm{s})\) \\
\(G_{P_v,\mathrm{out}}\) & Venous PVS downstream conductance &    \(50\) & \(\mu\mathrm{m}^3/(\mathrm{Pa}\cdot\mathrm{s})\) \\
\hline
\end{tabular}
\end{table}

\begin{table}[htbp]
\centering
\caption{Waveform, dynamic-gap, and tracer-transport parameters.}
\label{tab:waveform_gap_tracer_parameters}
\begin{tabular}{llll}
\hline
Parameter & Description & Baseline value & Unit \\
\hline
\(T_p\) & Vascular waveform period & \(10\) & s \\
\(\sigma_s\) & Width of symmetric vasodilation pulse & \(0.12\) & - \\
\(\phi_s\) & Center of symmetric vasodilation pulse & \(0.50\) & - \\
\(\tau_r\) & Rise time scale of asymmetric pulse & \(0.06\) & - \\
\(\tau_d\) & Decay time scale of asymmetric pulse & \(0.172\) & - \\
\(w_0\) & Baseline arterial gap factor & \(1.0\) & - \\
\(w_{\min}\) & Minimum gap factor & \(0.2\) & - \\
\(w_{\max}\) & Maximum gap factor & \(2.0\) & - \\
\(\tau_w\) & Gap relaxation time scale & \(1.0\) & s \\
\(g_{\mathrm{open}}\) & Compression-induced gap-opening strength & \(0.3\) & - \\
\(g_{\mathrm{volA}}\) & Endfoot-volume-to-gap coupling strength & \(10.0\) & - \\
\(S_{a,0}\) & Compression-source normalization & \(\max_{t\in[0,T_p]}|S_a(t)|\) & \(\mu\mathrm{m}^3/\mathrm{s}\) \\
\(\alpha_{\mathrm{AQP4}}\) & AQP4 scaling factor & \(1.0\) & - \\
\(D_{P_aE_a}\) & Effective tracer exchange between \(P_a\) and \(E_a\) & \(0\) & \(\mu\mathrm{m}^3/\mathrm{s}\) \\
\(D_{E_aE_m}\) & Effective tracer exchange between \(E_a\) and \(E_m\) & \(1\) & \(\mu\mathrm{m}^3/\mathrm{s}\) \\
\(D_{E_mE_v}\) & Effective tracer exchange between \(E_m\) and \(E_v\) & \(1\) & \(\mu\mathrm{m}^3/\mathrm{s}\) \\
\(D_{E_vP_v}\) & Effective tracer exchange between \(E_v\) and \(P_v\) & \(0.5\) & \(\mu\mathrm{m}^3/\mathrm{s}\) \\
\(c_{P_a,\mathrm{out}}\) & Arterial reservoir tracer concentration & 0& - \\
\(c_{P_v,\mathrm{out}}\) & Venous reservoir tracer concentration & \(0\) & - \\
\(T_{\mathrm{end}}\) & Simulation end time &  \(6000\) & s \\
\hline
\end{tabular}
\end{table}

The parameters in Tables~\ref{tab:model_parameters}-\ref{tab:waveform_gap_tracer_parameters} should be interpreted as effective reduced-model parameters rather than direct anatomical measurements. 
The reference volumes determine the relative storage capacities of the compartments, while the hydraulic conductances determine the characteristic exchange time scales between compartments. 
In particular, \(V_{E_m}^{\mathrm{ref}}>V_{E_a}^{\mathrm{ref}},V_{E_v}^{\mathrm{ref}}\) is chosen so that the middle ECS compartment acts as a spatial buffer. 
The downstream venous outflow conductance \(G_{P_v,\mathrm{out}}\) is varied in selected simulations to distinguish venous PVS storage from true downstream clearance.

\FloatBarrier

\section{Results}\label{sec:results}
In this section, we use the multicompartment model to determine how vascular
forcing is converted into directional perivascular transport and
downstream tracer clearance. We first compare cardiac, symmetric
slow-wave, and temporally asymmetric arteriolar forcing to distinguish
large bidirectional water exchange from rectified net PVS-ECS
transport. We then examine how dynamic inter-endfoot gap regulation
modifies this conversion through the phase-dependent coupling between
gap conductance and the PVS-ECS pressure difference, and how this
mechanism depends on the gap-response timescale. Next, we investigate
how the magnitude and perivascular polarization of functional
AQP4-associated conductance influence the dominant gap pathway
indirectly through endfoot water balance, compartment pressures, and
gap dynamics. Finally, we quantify the combined effects of
aging-associated reductions in vascular forcing, PVS mechanical
coupling, and PVS-facing AQP4 polarization. Throughout the analysis, we
distinguish instantaneous forward and backward water exchange, net
arterial PVS-to-ECS transport, tracer redistribution and storage within
the ECS-venous PVS network, and cumulative solute removal through the
downstream outlet.

We first introduce the quantities used to identify important effects. 
To distinguish bidirectional mixing from directional arterial
PVS-ECS transport, we define the cumulative forward and backward water
exchange through the arterial inter-endfoot gap as
\[
F_{\mathrm{forward}}(T)
=
\int_0^T
\left(Q_{P_aE_a}(t)\right)^+
\,dt
\quad \rm{and}\quad
F_{\mathrm{backward}}(T)
=
\int_0^T
\left(-Q_{P_aE_a}(t)\right)^+
\,dt.
\]
The cumulative net exchange is
\[
F_{\mathrm{net}}(T)
=
\int_0^T Q_{P_aE_a}(t)\,dt
=
F_{\mathrm{forward}}(T)
-
F_{\mathrm{backward}}(T).
\]

For the tracer dynamics, the total tracer mass remaining in the ECS is
\[
M_E^{\mathrm{tot}}(t)
=
M_{E_a}(t)+M_{E_m}(t)+M_{E_v}(t).
\]
The ECS remaining and depleted fractions are defined as
\[
R_E(T)
=
\frac{M_E^{\mathrm{tot}}(T)}
     {M_E^{\mathrm{tot}}(0)},
\qquad
C_E(T)
=
1-R_E(T).
\]
Here, \(C_E(T)\) measures tracer loss from the ECS compartments, but
does not necessarily represent complete removal from the modeled
system, because tracer may be temporarily stored in the arterial or
venous PVS.

True downstream removal is quantified by the cumulative venous output
\[
\mathcal O_v(T)
=
\int_0^T
\left(J_{P_v,\mathrm{out}}(t)\right)^+
\,dt.
\]
The venous output fraction, normalized by the initial ECS tracer mass,
is
\[
F_v(T)
=
\frac{\mathcal O_v(T)}
     {M_E^{\mathrm{tot}}(0)}.
\]

To distinguish venous PVS storage from efficient downstream drainage,
we define the venous outlet efficiency
\[
\eta_v(T)
=
\frac{\mathcal O_v(T)}
     {M_{P_v}(T)+\mathcal O_v(T)}.
\]
A small value of \(\eta_v\) indicates that tracer has reached the venous
PVS but remains stored there rather than being rapidly removed through
the downstream outlet.

Finally, to identify transport bottlenecks along the
arterial-ECS-venous pathway, we compute the cumulative internal tracer
fluxes between $k_{\rm th}$ and $j_{\rm th}$ compartments
\[
\mathcal J_{kl}(T)
=
\int_0^T J_{kl}(t)\,dt,
\]
Together, these quantities distinguish arterial-side redistribution,
transport through the ECS chain, delivery to the venous PVS, temporary
PVS storage, and final downstream clearance.

All compartment pressures are measured relative to a common
physiological reference pressure. Therefore,
\[
p_{P_a}(0)=p_A(0)=p_{E_a}(0)=p_{E_m}(0)
=p_{E_v}(0)=p_{P_v}(0)=0
\]
represents an equilibrated reference state and does not imply zero
absolute physiological pressure. Because the fluxes depend only on
pressure differences, adding a common constant pressure to all
compartments does not affect the model dynamics.

\subsection{Waveform-dependent water exchange under fixed-gap conditions}
\label{sec:fixed_gap_water}

We first examine how vascular waveform shape affects pressure-driven
water exchange in the reduced arterial-ECS-chain-venous model. The
arterial inter-endfoot gap is held fixed,
\[
w_a(t)\equiv w_0,
\]
so that the comparison isolates the effect of the prescribed vascular
forcing. We consider a cardiac-like zero-mean oscillation, a
time-symmetric vasodilation waveform, and a fast-rise slow-decay
asymmetric vasodilation waveform. All cases use the healthy polarized
AQP4 state,
\(
\rho_{\rm AQP4}=0.9,
\)
corresponding to
\(
L_{P_aA}^{\rm eff}=0.05,
L_{AE_a}^{\rm eff}=0.01.
\)

\begin{figure}[!htp]
    \centering
   \subfigure[] {\includegraphics [width=0.6\linewidth]{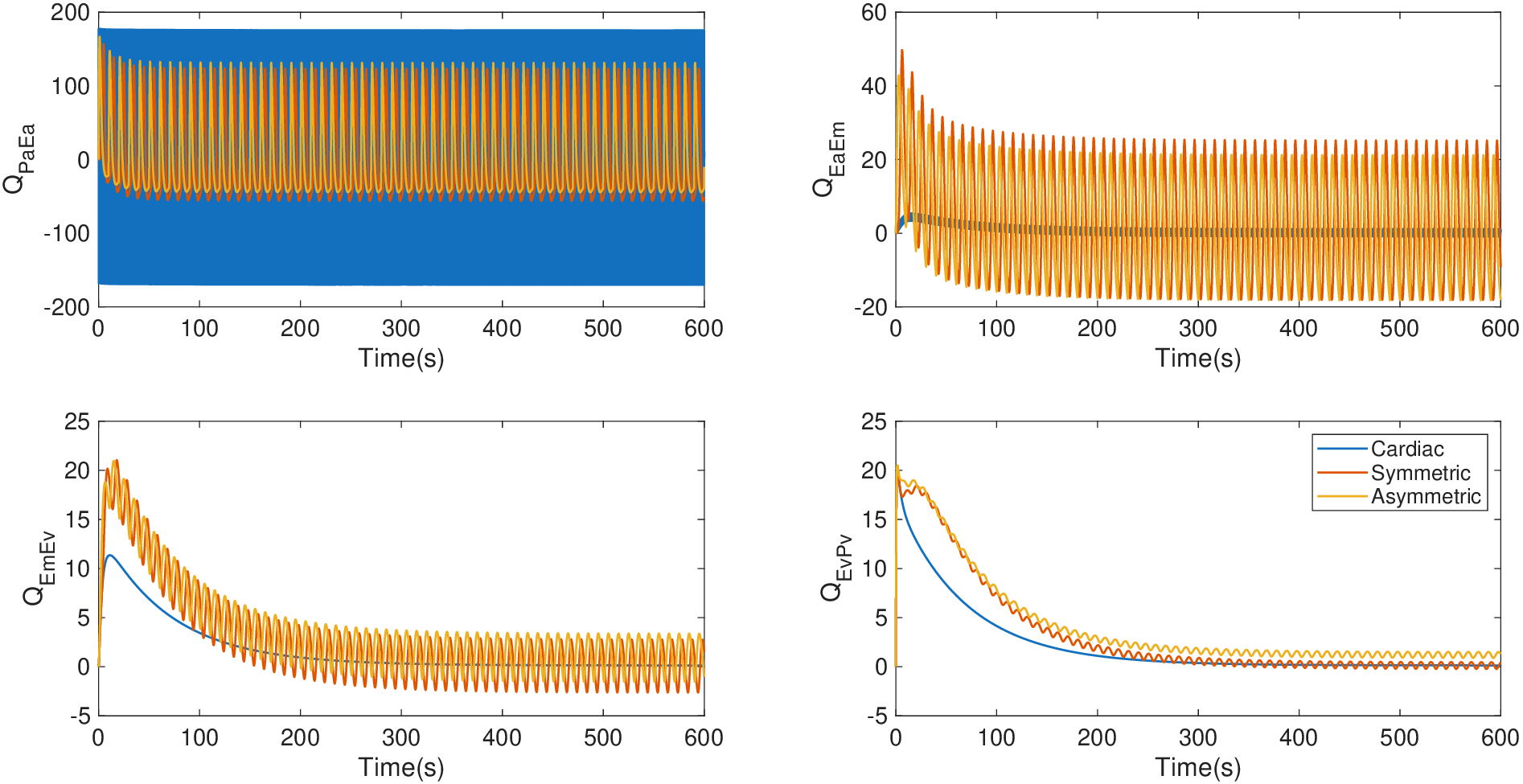}}
    \subfigure[]{\includegraphics[width=0.6\linewidth]{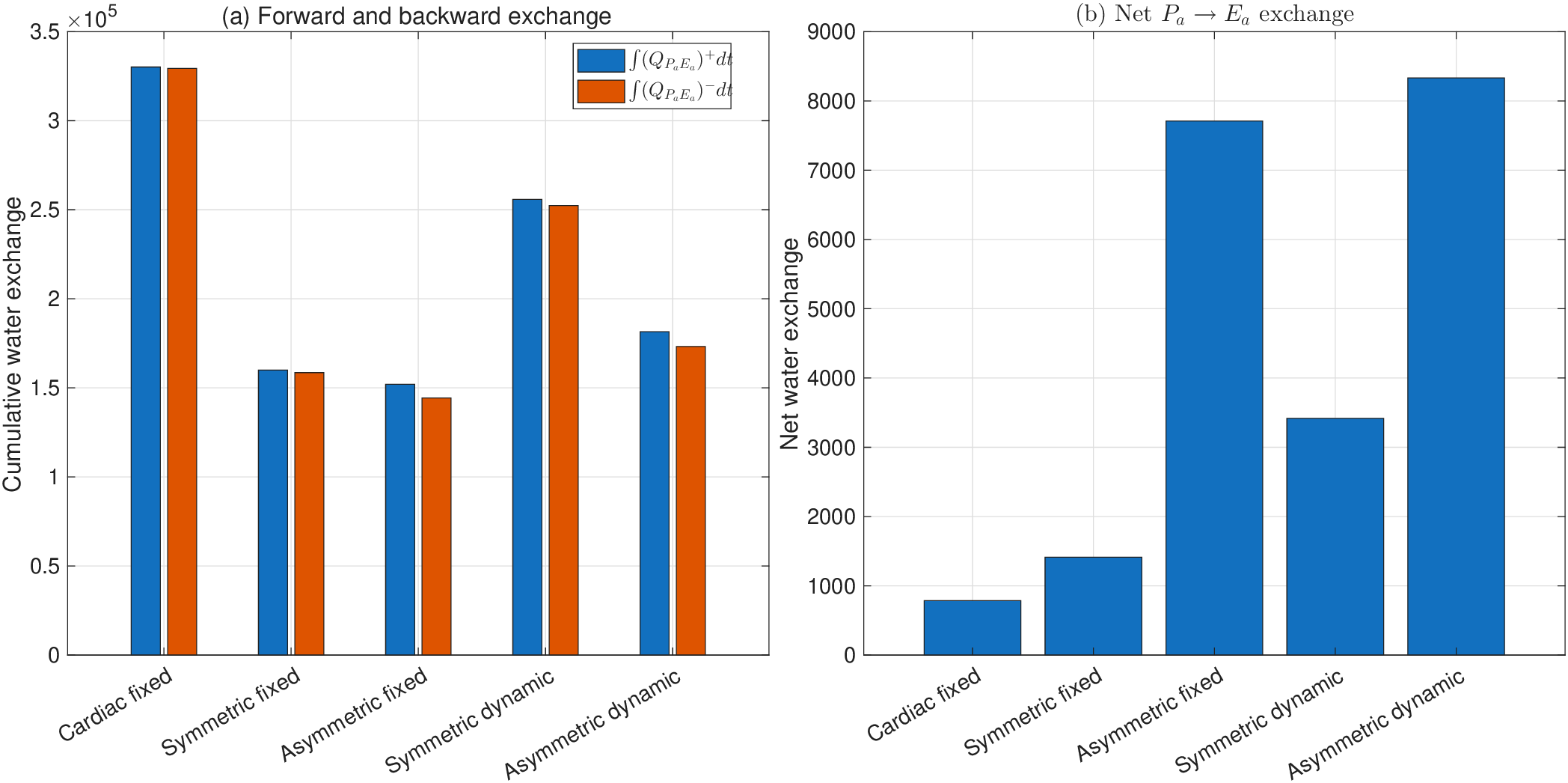}}
    \caption{ Waveform-dependent water exchange under fixed-gap conditions.
    (a) Water fluxes along the arterial-ECS-chain-venous pathway.
    (b) Cumulative forward, backward, and net arterial PVS-to-ECS
    exchange. Cardiac-like forcing produces the largest bidirectional
    exchange but weak net transport, whereas asymmetric vasodilation
    produces the strongest directional transport by suppressing
    recovery-phase backflow.}
    \label{fig:fix_gap_results}
\end{figure}

Figure~\ref{fig:fix_gap_results} shows the water fluxes over the
long-time repeated-waveform simulation. 
The vasodilation waveforms generate smaller total bidirectional exchange
than the cardiac waveform but substantially larger net transport.
In particular, the asymmetric fast-rise slow-decay waveform produces
approximately \(5.5\) times the net arterial PVS-to-ECS transport of the
symmetric waveform. This enhancement arises primarily from reduced
recovery-phase backflow rather than from an increase in cumulative
forward exchange. Thus, temporal waveform asymmetry promotes
directional transport by weakening forward-backward cancellation. Its net transport is also approximately \(9.8\) times that generated by
the cardiac-like waveform.

The downstream fluxes \(Q_{E_aE_m}\), \(Q_{E_mE_v}\), and
\(Q_{E_vP_v}\) are smoother than the local arterial gap flux because the
ECS compartments provide finite storage and hydraulic buffering. In
particular, the middle ECS compartment \(E_m\) represents a lumped bulk
parenchymal reservoir that attenuates rapid oscillations and introduces
a propagation delay between arterial-side exchange and venous-side
drainage. The asymmetric waveform maintains larger downstream fluxes,
showing that rectification generated at the arterial interface can
propagate through the ECS chain toward the venous PVS.

These findings are qualitatively consistent with poroelastic
simulations showing that temporally asymmetric functional hyperemia can
promote directional PVS-to-ECS transport, whereas periodic wall motion
primarily produces bidirectional exchange with limited net displacement
\cite{Kedarasetti2022}. Although the present model is spatially lumped,
it reproduces the same waveform-dependent rectification mechanism:
temporal asymmetry suppresses recovery-phase backflow and thereby
enhances net transport.

Under fixed-gap conditions, the gap-mediated flux is substantially
larger than the direct PVS-facing endfoot flux. The peak-flux ratios are
\[
\frac{\max |Q_{P_aE_a}|}{\max |Q_{P_aA}|}
=
20.0,\quad 19.7,\quad 19.8
\]
for the cardiac, symmetric, and asymmetric waveforms, respectively.
The corresponding values are summarized in
Table~\ref{tab:fixed_gap_water_flux}. Thus,
\[
|Q_{P_aA}|\ll |Q_{P_aE_a}|,
\]
and hydrostatic exchange is dominated by the inter-endfoot gap pathway.

The net PVS-facing endfoot flux is negative in all three cases,
whereas the net gap-mediated flux is positive. The endfoot pathway
therefore does not carry the dominant net arterial PVS-to-ECS transport
under the present conditions. Its importance instead arises through
its effect on the coupled pressure-volume balance, as examined below.
The precise peak-flux ratio reflects the prescribed effective
conductances and should not be interpreted as a quantitative
microscale prediction, but the pathway ordering is consistent with
geometry-resolved models of the gliovascular interface
\cite{Causemann2026}. 

\begin{table}[htbp]
\centering
\caption{Comparison of gap-mediated and PVS-facing endfoot water fluxes
under fixed-gap conditions.}
\label{tab:fixed_gap_water_flux}
\begin{tabular}{lccccc}
\hline
Case
& $\max |Q_{P_aE_a}|$
& $\max |Q_{P_aA}|$
& Ratio
& $F^{\rm net}_{P_aE_a}$
& $F^{\rm net}_{P_aA}$ \\
\hline
Cardiac fixed
& 177.6
& 8.882
& 20.0
& 786.6
& $-652.5$ \\

Symmetric fixed
& 156.7
& 7.940
& 19.7
& 1412.9
& $-630.5$ \\

Asymmetric fixed
& 166.7
& 8.402
& 19.8
& 7710
& $-398.2$ \\
\hline
\end{tabular}
\end{table}

\subsection{Waveform-dependent ECS-chain tracer clearance}
\label{sec:waveform_clearance}

We next examine whether waveform-dependent water exchange produces
effective tracer clearance through the arterial-ECS-chain-venous
pathway. The simulations use the uniform ECS-loading protocol
\[
c_{E_a}(0)=c_{E_m}(0)=c_{E_v}(0)=1,
\qquad
c_{P_a}(0)=c_{P_v}(0)=0,
\]
with tracer-free external reservoirs. The arterial outlet conductance
is set to zero, so tracer transported from \(E_a\) into the arterial PVS
is stored in \(P_a\) rather than removed. This separates arterial-side
redistribution, venous PVS storage, and true downstream venous output.

The fixed-gap results in
Table~\ref{tab:waveform_clearance_summary} show that waveform shape
strongly affects downstream clearance. Despite producing the largest
bidirectional water exchange, the cardiac waveform gives the smallest
cumulative venous output,
\(
O_v(T)=136.5.
\)
The symmetric waveform increases this value to
\(
O_v(T)=267.0,
\)
whereas the asymmetric waveform produces
\(
O_v(T)=2036.7.
\)
Thus, under fixed-gap conditions, asymmetric forcing generates
approximately \(7.6\) times the venous output of symmetric forcing and
approximately \(14.9\) times that of cardiac forcing.

The corresponding ECS depleted fractions are
\(
12.66\%, 14.21\%, 17.90\%
\)
for the cardiac, symmetric, and asymmetric waveforms, respectively.
However, the venous output fractions remain much smaller:
\(
0.123\%, 0.241\%, 1.84\%.
\)
Therefore, loss of tracer from the ECS does not imply complete
downstream clearance. A substantial fraction is redistributed into the
arterial and venous PVS compartments or remains stored within the
transport chain.

Venous PVS storage also remains substantial. At the final time,
\(
M_{P_v}(T)
=
4237,  4929, 8692
\)
for the cardiac, symmetric, and asymmetric fixed-gap cases,
respectively. The corresponding venous outlet efficiencies are
\(
3.12\%, 5.14\%, 18.98\%.
\)
The asymmetric waveform therefore improves not only transport toward
the venous PVS, but also subsequent downstream removal.

The cumulative arterial-side tracer flux
\(J_{P_aE_a}^{\rm cum}\) remains negative for all waveforms, indicating
persistent tracer redistribution from \(E_a\) toward the arterial PVS.
In contrast, the downstream tracer fluxes through
\(E_m\), \(E_v\), and \(P_v\) increase markedly under asymmetric
forcing. These results show that temporal asymmetry promotes tracer
propagation through the full ECS-chain-venous pathway rather than
merely increasing local arterial exchange.

\begin{figure}
 \centering
   \subfigure[] {\includegraphics [width=0.45\linewidth]{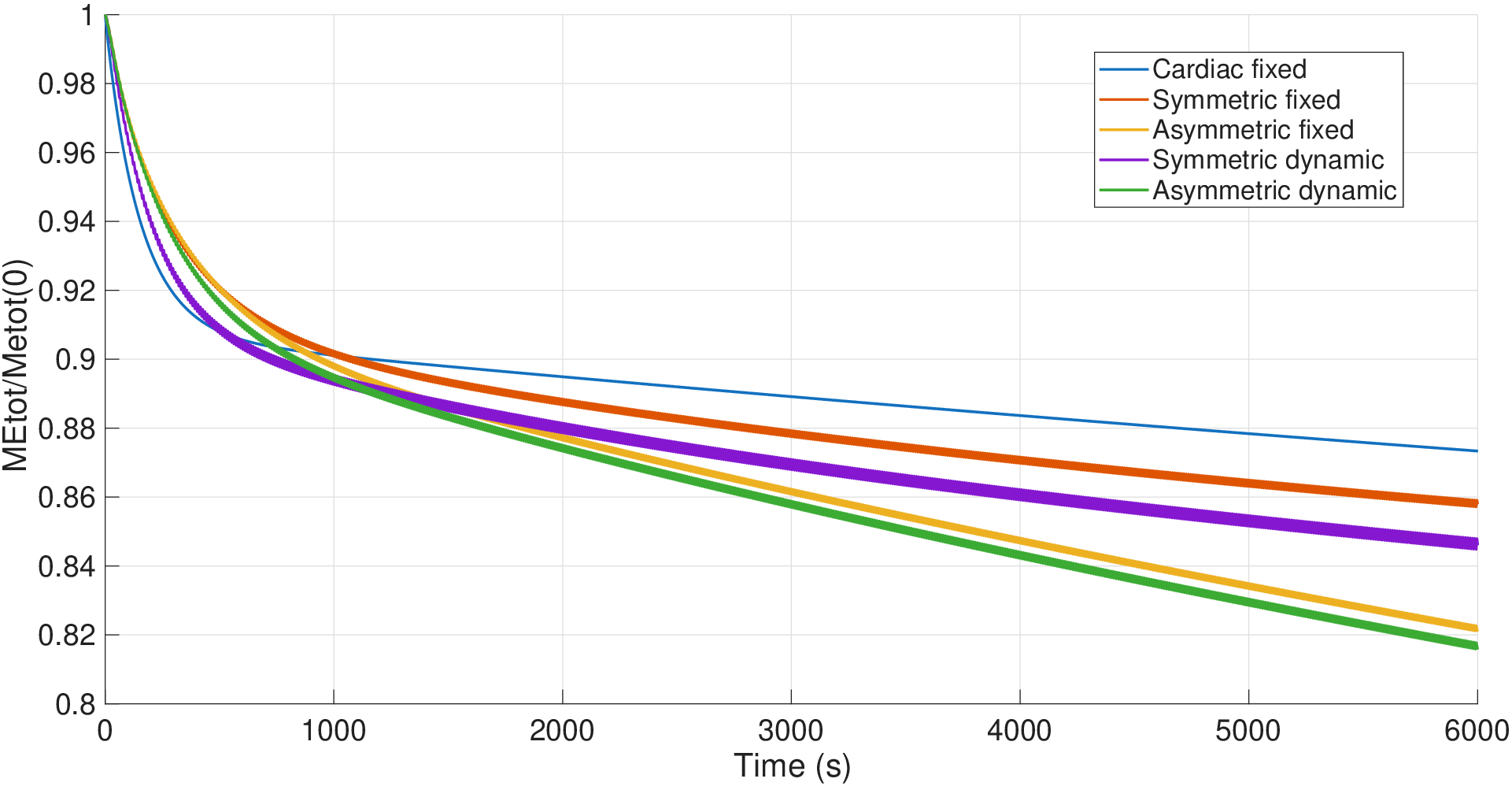}}
    \subfigure[]{\includegraphics[width=0.45\linewidth]{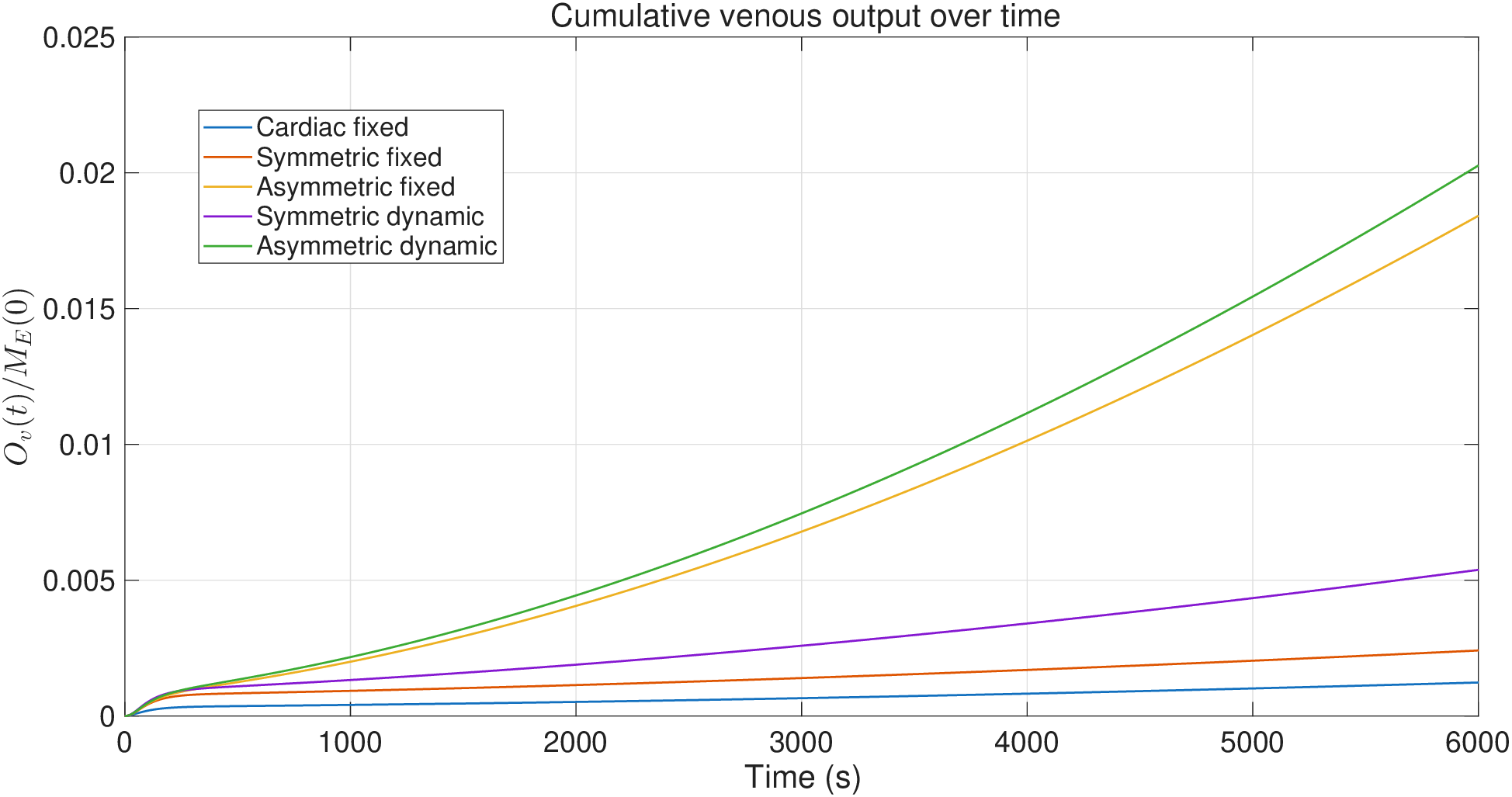}}
    \subfigure[]{\includegraphics[width=0.45\linewidth]{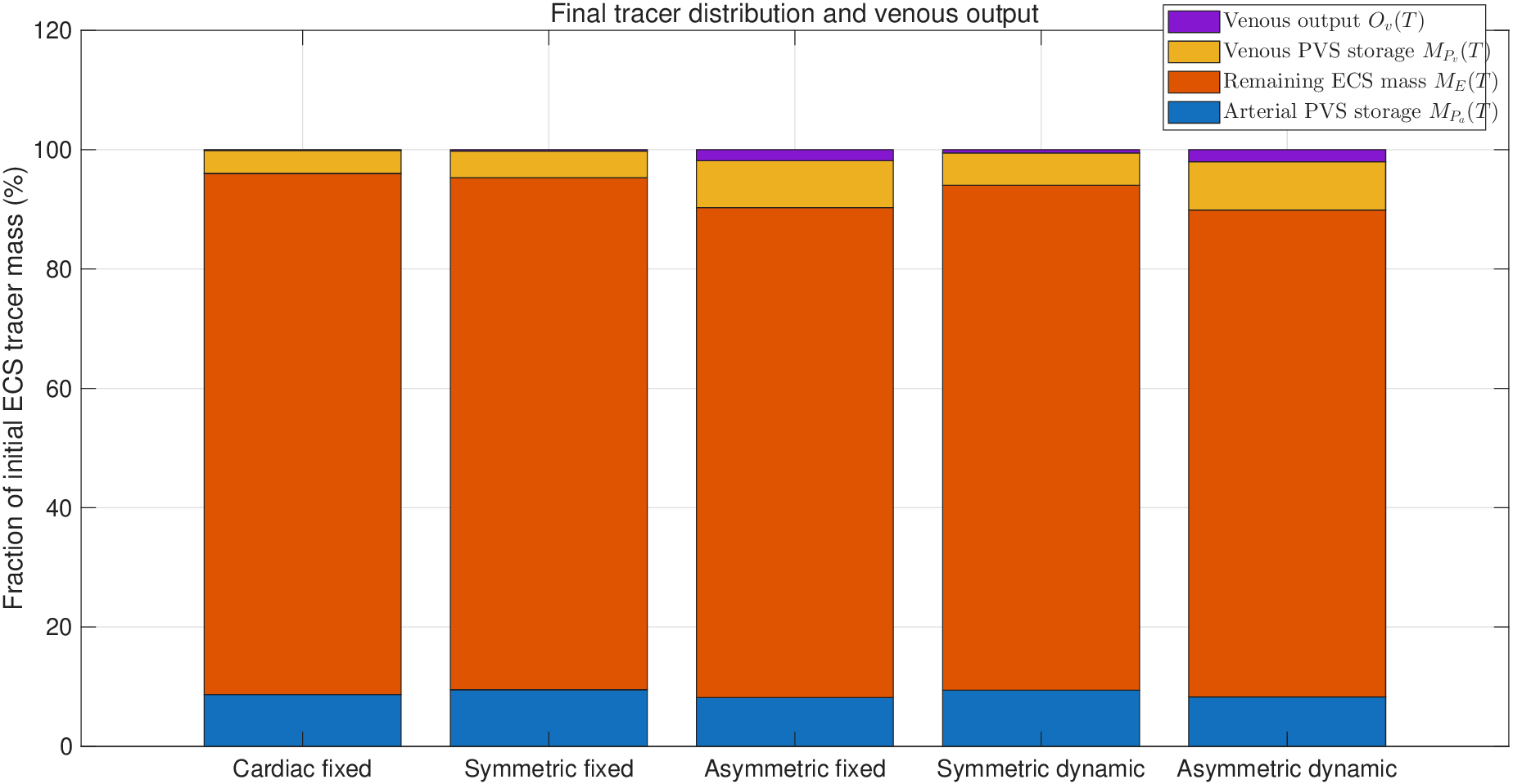}}
    \subfigure[]{\includegraphics[width=0.45\linewidth]{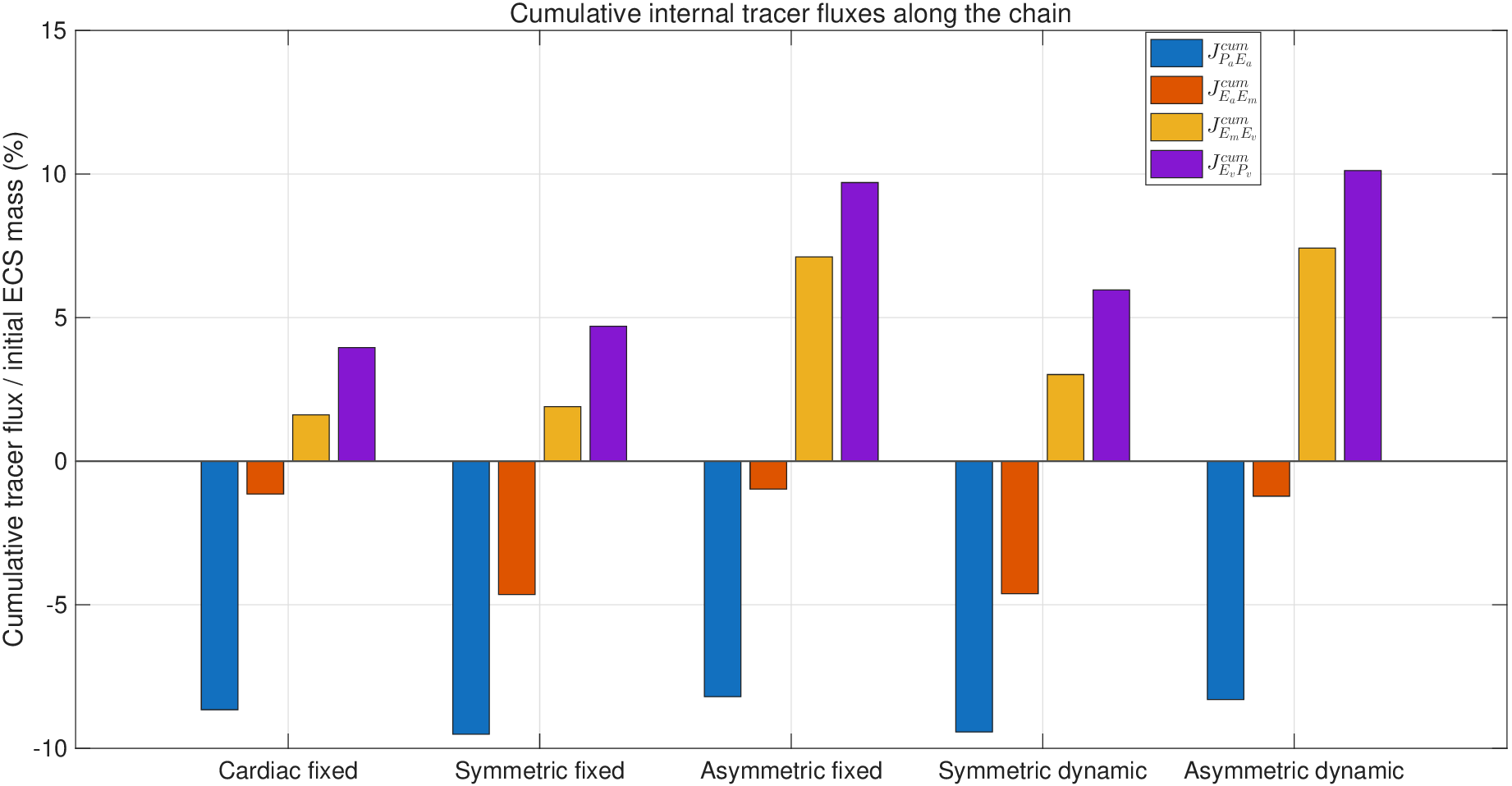}}
\caption{Long-time ECS-chain tracer transport.
    (a) Normalized total ECS tracer mass.
    (b) Normalized cumulative venous output.
    (c) Final tracer distribution.
    (d) Cumulative internal tracer fluxes.
    Asymmetric forcing produces the strongest downstream clearance,
    while dynamic gap regulation provides its largest enhancement under
    symmetric forcing.}
\label{fig:waveform_clearance_summary}
\end{figure}

Figure~\ref{fig:waveform_clearance_summary} also includes the
corresponding dynamic-gap results. Dynamic gap regulation provides a
large additional enhancement under symmetric forcing but only a modest
increase under asymmetric forcing. This waveform-dependent
amplification is analyzed in the following subsection.

\begin{table}[htbp]
\centering
\caption{Long-time ECS-chain tracer clearance under fixed-gap and dynamic-gap waveform conditions.}
\label{tab:waveform_clearance_summary}
\begin{tabular}{lccccc}
\hline
Case
& $C_E(T)$
& $O_v(T)$
& $F_v(T)$
& $M_{P_v}(T)$
& $\eta_v(T)$ \\
\hline
Cardiac fixed
& $12.66\%$
& 136.54
& $0.123\%$
& 4237
& $3.122\%$ \\
Symmetric fixed
& $14.21\%$
& 266.97
& $0.241\%$
& 4929
& $5.138\%$ \\
Asymmetric fixed
& $17.90\%$
& 2036.69
& $1.84\%$
& 8692
& $18.984\%$ \\
Symmetric dynamic
& $15.40\%$
& 595.20
& $0.538\%$
& 5998
& $9.027\%$ \\
Asymmetric dynamic
& $18.42\%$
& 2241.59
& $2.03\%$
& 8953
& $20.025\%$ \\
\hline
\end{tabular}
\end{table}

\subsection{Dynamic inter-endfoot gap regulation amplifies waveform-driven transport}

The fixed-gap simulations show that waveform shape by itself strongly
affects downstream tracer transport. We next allow the arterial
inter-endfoot gap to respond dynamically to vascular compression and
endfoot-volume changes. All simulations in this comparison use the
healthy polarized AQP4 state,
\(\rho_{\mathrm{AQP4}}=0.9\).

Figure~\ref{fig:waveform_clearance_summary} compares fixed- and
dynamic-gap simulations over the long-time clearance window. Dynamic
gap regulation increases net arterial PVS-to-ECS water transport for
both vasodilation waveforms, although the magnitude of the enhancement
is strongly waveform dependent. Under symmetric forcing,
\(
F_{P_aE_a}^{\mathrm{net}}
:
1412.9 \longrightarrow 3416.6,
\)
corresponding to an increase of approximately \(142\%\). Under
asymmetric forcing, the increase is more modest,
\(
F_{P_aE_a}^{\mathrm{net}}
:
7710 \longrightarrow 8332,
\)
or approximately \(8\%\).

The different transport enhancements reflect the corresponding gap
responses. For the symmetric waveform,
\(
w_a(t)\in[0.9772,1.3085],
\qquad
\overline{w_a}=1.1466.
\)
Because
\(
G_{a,\mathrm{gap}}\propto w_a^3,
\)
the maximum gap conductance reaches approximately \(2.24\) times its
baseline value. In contrast, under asymmetric forcing,
\[
w_a(t)\in[0.9662,1.115],
\qquad
\overline{w_a}=1.056,
\]
and the maximum conductance is only approximately \(1.39\) times the
baseline value. Dynamic regulation therefore produces substantially
stronger conductance amplification under symmetric forcing.

The tracer-clearance metrics exhibit the same waveform dependence. For
the symmetric waveform, cumulative venous output increases from
\(
O_v(T)=266.97
\)
under fixed-gap conditions to
\(
O_v(T)=595.20
\)
with dynamic gap regulation, corresponding to an increase of
approximately \(123\%\). For the asymmetric waveform, the increase is
from
\(
O_v(T)=2036.69
\)
to
\(
O_v(T)=2241.59,
\)
or approximately \(10\%\). Thus, dynamic gap regulation provides its
largest relative benefit when directional rectification by the
waveform itself is weak. Under strongly asymmetric forcing, suppression
of recovery-phase backflow itself generates substantial net transport,
leaving a smaller marginal role for conductance modulation.

The internal tracer fluxes further show the same effects.
Dynamic gap regulation enhances transport through the
ECS-chain-venous pathway, particularly under symmetric forcing.
However, the cumulative arterial-side tracer flux
\(J_{P_aE_a}^{\mathrm{cum}}\) remains negative in all cases, indicating
that tracer transfer from \(E_a\) back toward the arterial PVS persists
even when the gap is dynamic. Dynamic gap regulation therefore does not
eliminate arterial-side tracer back-exchange. Instead, it increases
gap-mediated water exchange and improves downstream propagation when
gap opening is favorably aligned with the PVS-ECS pressure difference.

Overall, dynamic inter-endfoot gap regulation acts as a
  conductance amplifier. Its relative contribution is
largest under symmetric slow vasomotion. The effects of AQP4
abundance and polarization are examined separately below.
\subsection{Mechanistic  decomposition analysis of dynamic gap regulation}

\begin{figure}[!htp]
    \centering
    \includegraphics[width=0.75\linewidth]{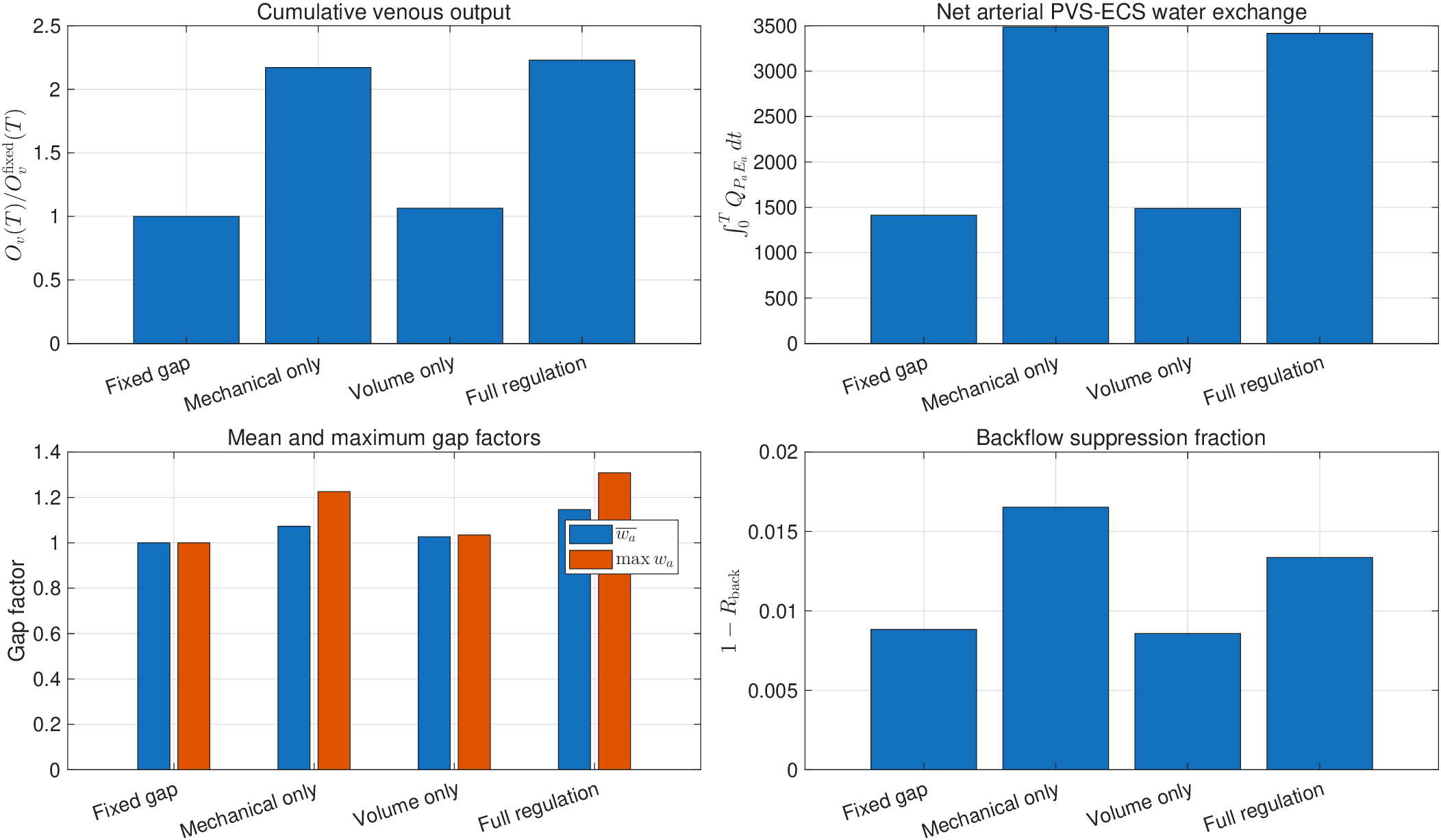}
    \caption{Mechanism ablation of dynamic arterial inter-endfoot gap
regulation under symmetric slow-vasomotion forcing. The fixed-gap case
is compared with mechanical regulation only, endfoot-volume regulation
only, and the full dynamic-gap model. Vascularly induced mechanical
opening provides the dominant enhancement of cumulative venous tracer
output and net arterial PVS-ECS water exchange. Endfoot-volume feedback
alone has a smaller effect by itself. When combined with mechanical regulation, further increases occur in cumulative venous output and the maximum gap factor. Despite large bidirectional water exchange in all cases,
dynamic mechanical regulation increases the small net transport
fraction.}
    \label{fig:gap_ablation}
\end{figure}

Figure~\ref{fig:gap_ablation} compares the fixed-gap model with
mechanical regulation only (\(g_{\rm vol}=0\)), endfoot-volume
regulation only (\(g_{\rm open}=0\)), and the full dynamic-gap model.
The corresponding quantitative metrics are summarized in
Table~\ref{tab:gap_ablation}.

Under symmetric slow-vasomotion forcing, vascularly induced mechanical
opening is the dominant source of dynamic-gap enhancement. Relative to
the fixed-gap case, mechanical regulation alone increases cumulative
venous tracer output from \(266.97\) to \(579.62\), corresponding to an
increase of approximately \(117\%\). Endfoot-volume regulation alone
produces only a modest increase to \(284.34\), or approximately \(6.5\%\).
The full dynamic-gap model yields \(O_v(T)=595.20\), corresponding to an
increase of approximately \(123\%\) relative to the fixed-gap case.
Thus, endfoot-volume feedback provides only a small additional increase
in downstream venous output once mechanically induced opening is
present.

The gap statistics further distinguish the two mechanisms. Mechanical
regulation produces only gap widening, with
\[
w_a\in[1.0000,1.2256],
\qquad
\overline{w_a}=1.0735.
\]
Volume regulation alone produces both narrowing and widening,
\[
w_a\in[0.9610,1.0351],
\qquad
\overline{w_a}=1.0264,
\]
reflecting alternating endfoot swelling and shrinkage during different
phases of the waveform. In the full model,
\[
w_a\in[0.9772,1.3085],
\qquad
\overline{w_a}=1.1466.
\]
These results show that  larger mean and peak gap
opening do not necessarily produce larger net arterial
PVS-to-ECS water transport. Mechanical regulation alone gives
\(
J_{P_aE_a}^{\rm net}=3487.9,
\)
whereas the full model gives the slightly smaller value
\(
J_{P_aE_a}^{\rm net}=3416.6,
\)
despite its larger mean and maximum gap factors. Net transport depends not only on the magnitude of gap
opening, but also on the temporal alignment between gap conductance and
the PVS-ECS pressure difference.

For cumulative venous output, the combined effect is approximately
additive: the separate mechanical and volume contributions added together predict an
output close to that of the full model. In contrast, this approximate
additivity does not hold for net gap-mediated water exchange. Thus,
mechanical opening provides the dominant enhancement, while
endfoot-volume feedback mainly modulates the amplitude and timing of the
gap response.

\begin{table}[htbp]
\centering
\small
\caption{Separate contributions of mechanical and endfoot-volume regulation under symmetric slow-vasomotion forcing. Here,
\(\eta_{\rm net}=1-R_{\rm back}\) denotes the net directional fraction
of the cumulative forward exchange.}
\label{tab:gap_ablation}
\begin{tabular}{lcccccc}
\hline
Case
& \(O_v(T)\)
& \(J_{P_aE_a}^{\rm net}\)
& \(\eta_{\rm net}\) (\%)
& \(\overline{w_a}\)
& \(\min w_a\)
& \(\max w_a\) \\
\hline
Fixed gap
& 266.97
& 1412.9
& 0.883
& 1.0000
& 1.0000
& 1.0000 \\

Mechanical only
& 579.62
& 3487.9
& 1.653
& 1.0735
& 1.0000
& 1.2256 \\

Volume only
& 284.34
& 1488.8
& 0.858
& 1.0264
& 0.9610
& 1.0351 \\

Full regulation
& 595.20
& 3416.6
& 1.336
& 1.1466
& 0.9772
& 1.3085 \\
\hline
\end{tabular}
\end{table}

\begin{figure}
 \centering
    \subfigure[]{\includegraphics[width=0.45\linewidth]{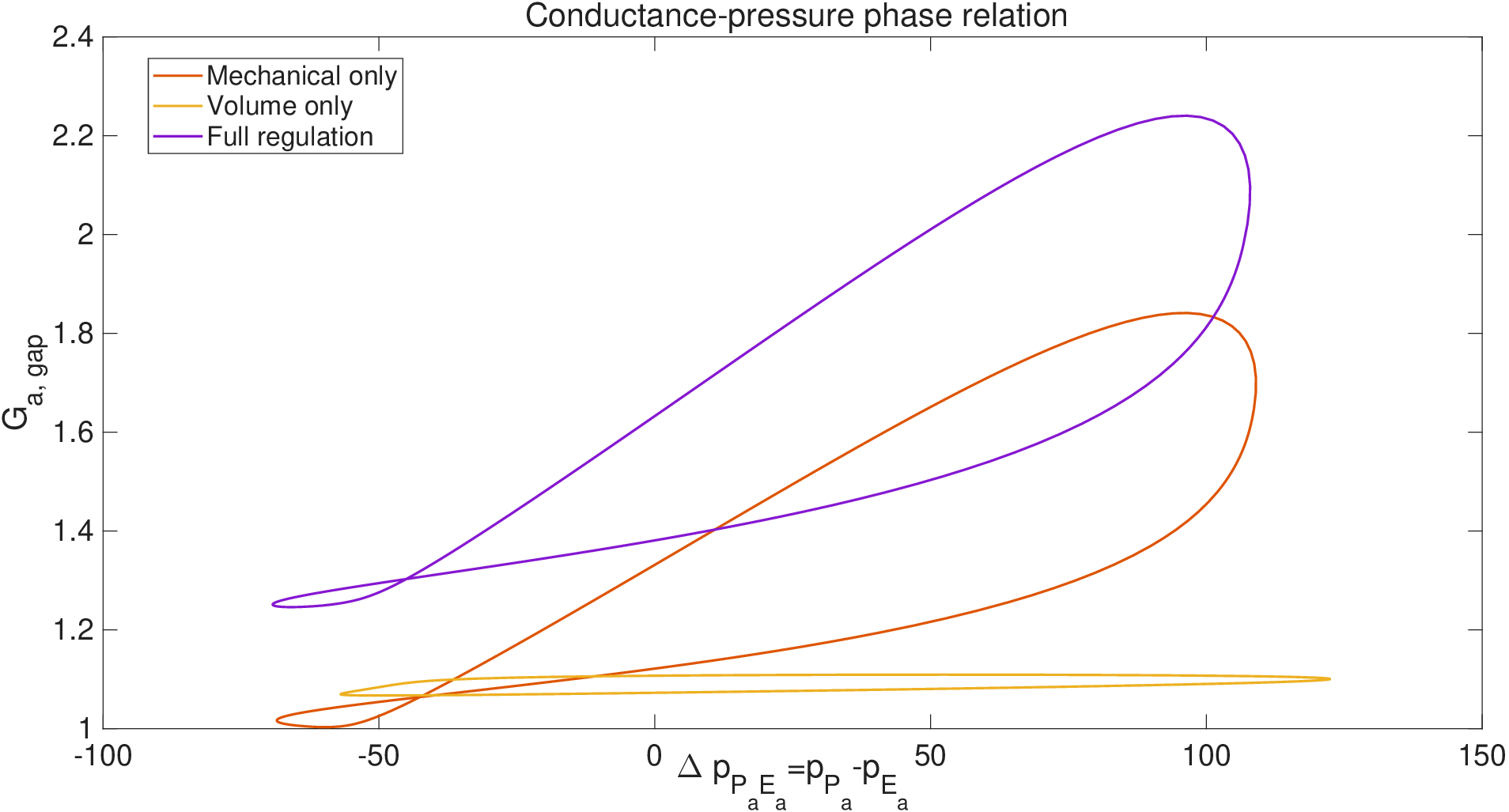}}
    \subfigure[]{\includegraphics[width=0.45\linewidth]{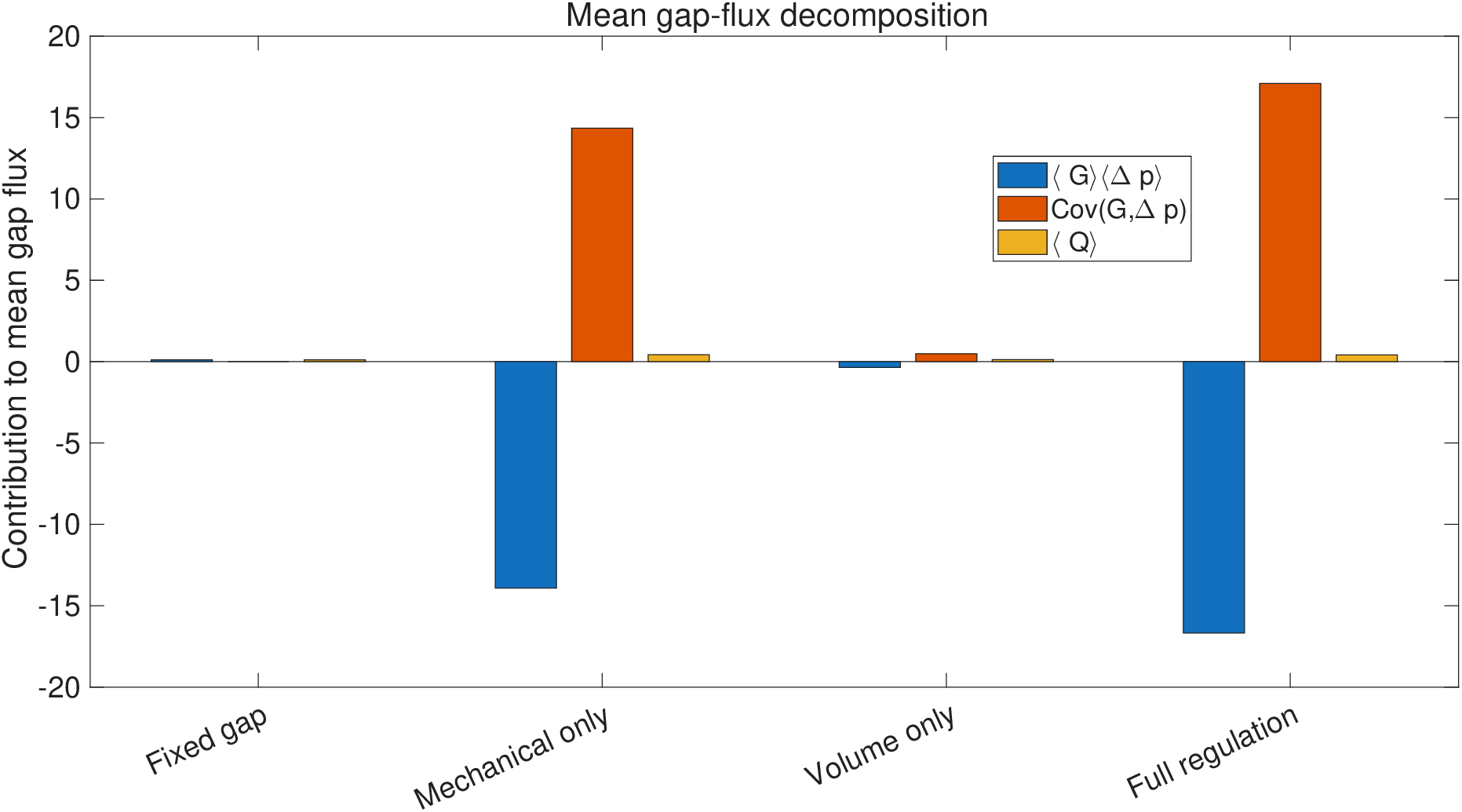}}
       \subfigure[] {\includegraphics [width=0.65\linewidth]{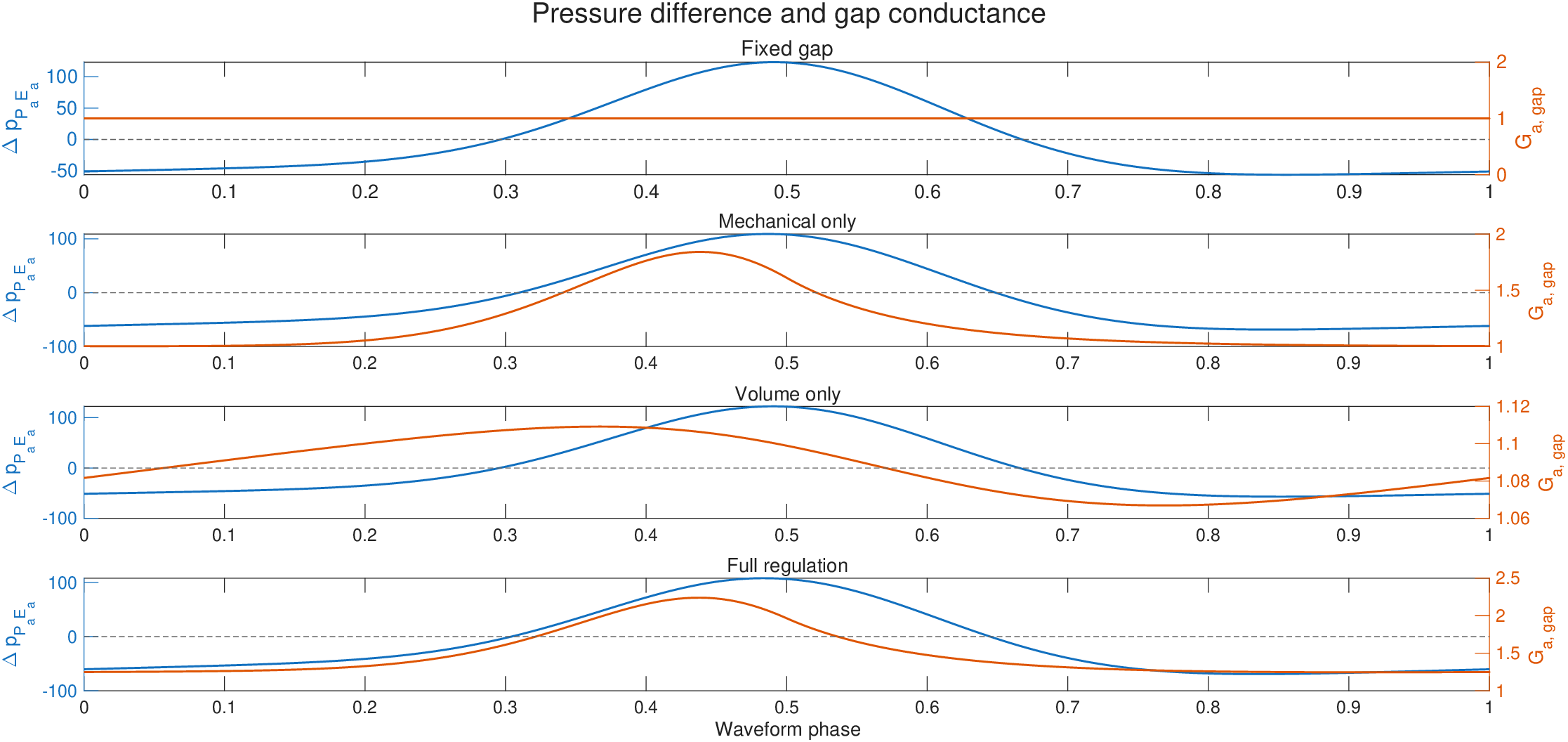}}
\caption{
(a) Conductance-pressure phase relation for the four differnet cases.
(b) Decomposition of the cycle-averaged gap flux.
Mechanical gap regulation generates a positive
conductance-pressure covariance that overcomes the adverse mean-pressure
contribution and produces positive net transport. (c) Temporal evolution of the arterial PVS-ECS pressure difference
\(\Delta p_{P_aE_a}=p_{P_a}-p_{E_a}\) and gap conductance
\(G_{a,\mathrm{gap}}\) over the last periodic cycle.}
\label{fig:pressurediff-gap-ralation}
\end{figure}

To clarify the mechanism underlying the ablation results, we examined
the temporal and phase relationships between the arterial PVS-ECS
pressure difference and the gap conductance over the last periodic
cycle in Fig.\ref{fig:pressurediff-gap-ralation}. Since
\[
Q_{P_aE_a}(t)
=
G_{a,\mathrm{gap}}(t) 
 (p_{P_a}-p_{E_a}),
\]
the sign of the instantaneous flux is determined by the pressure
difference, whereas its magnitude is modulated by the time-dependent
gap conductance.

For the mechanical-only and full-regulation cases, the conductance
increases predominantly during the positive-pressure, forward-flow
phase and decreases during the negative-pressure, reverse-flow phase.
This phase-dependent relationship is also evident from the
conductance-pressure loops: larger values of
\(G_{a,\mathrm{gap}}\) occur mainly on the
\(\Delta p_{P_aE_a}>0\) branch, whereas the conductance is reduced over
much of the \(\Delta p_{P_aE_a}<0\) branch. In contrast, the fixed-gap
case has constant conductance, while the volume-only case exhibits only
a small conductance variation and a much narrower phase loop.

To clarify how the time-dependent gap conductance contributes to net transport, 
we decompose both the gap conductance and the arterial PVS-ECS pressure difference 
into their cycle-averaged and fluctuating components:
\(
G_{a,\mathrm{gap}}(t)
=
\left\langle G_{a,\mathrm{gap}}\right\rangle
+
G'_{a,\mathrm{gap}}(t),
\)
and
\(
\Delta p_{P_aE_a}(t)
=
\left\langle \Delta p_{P_aE_a}\right\rangle
+
\Delta p'_{P_aE_a}(t),
\)
where
\(
\Delta p_{P_aE_a}=p_{P_a}-p_{E_a},
\)
and the prime denotes the deviation from the cycle average. The cycle-averaged flux can be written as
\[
\left\langle Q_{P_aE_a}\right\rangle
=
\left\langle G_{a,\mathrm{gap}}\right\rangle
\left\langle \Delta p_{P_aE_a}\right\rangle
+
\left\langle
G'_{a,\mathrm{gap}}
\Delta p'_{P_aE_a}
\right\rangle .
\]

The first term represents the contribution from the mean gap conductance and the
mean pressure difference. The second term, denoted by \(cov(G_{a,\mathrm{gap}},\Delta p_{P_aE_a})\), represents the contribution arising
from the temporal alignment between the time-dependent gap conductance and the
pressure difference over the vascular cycle. In particular, this term becomes
positive when larger gap conductance occurs preferentially during the
positive-pressure, forward-flow phase and smaller conductance occurs during the
negative-pressure, reverse-flow phase.
For the mechanically regulated cases, the unweighted mean pressure
difference is negative, but the positive conductance-pressure
covariance is slightly larger in magnitude and the mean flux is positive. Thus, the regulation of the dynamic gap acts as a hydraulic rectifier:
the dynamic gap   preferentially increases conductance during forward flow and reduces
conductance during reverse flow, thereby enhancing net arterial PVS-to-ECS transport by weakening forward-backward cancellation.

The full-regulation case produces a larger mean and peak
conductance than the mechanical-only case. Its high-conductance phase
extends farther into the declining and reverse-pressure portions of the
cycle. It therefore generates both a larger positive covariance and a
more adverse mean-pressure contribution. The resulting mean flux is
slightly smaller than in the mechanical-only case, demonstrating that
net transport depends on the timing of gap opening, rather than on gap
magnitude alone.

These results show that dynamic gap regulation enhances net transport
through the temporal alignment of gap conductance with the driving
pressure difference. In particular, the magnitude of gap opening alone
is insufficient to determine transport, because conductance must
increase during the forward-pressure phase and decrease during the
reverse-pressure interaction (in time) to produce effective rectification. This
observation motivates examining the gap response timescale
\(\tau_w\), which controls the delay between the target gap opening and
the actual gap dynamics. We therefore next investigate how the ratio
\(\tau_w/T_{\rm wave}\) affects conductance-pressure phase coupling and
the resulting clearance.

\subsection{Sensitivity to the arterial gap response time}

\begin{figure}[!htb]
    \centering
    \includegraphics[width=0.75\linewidth]{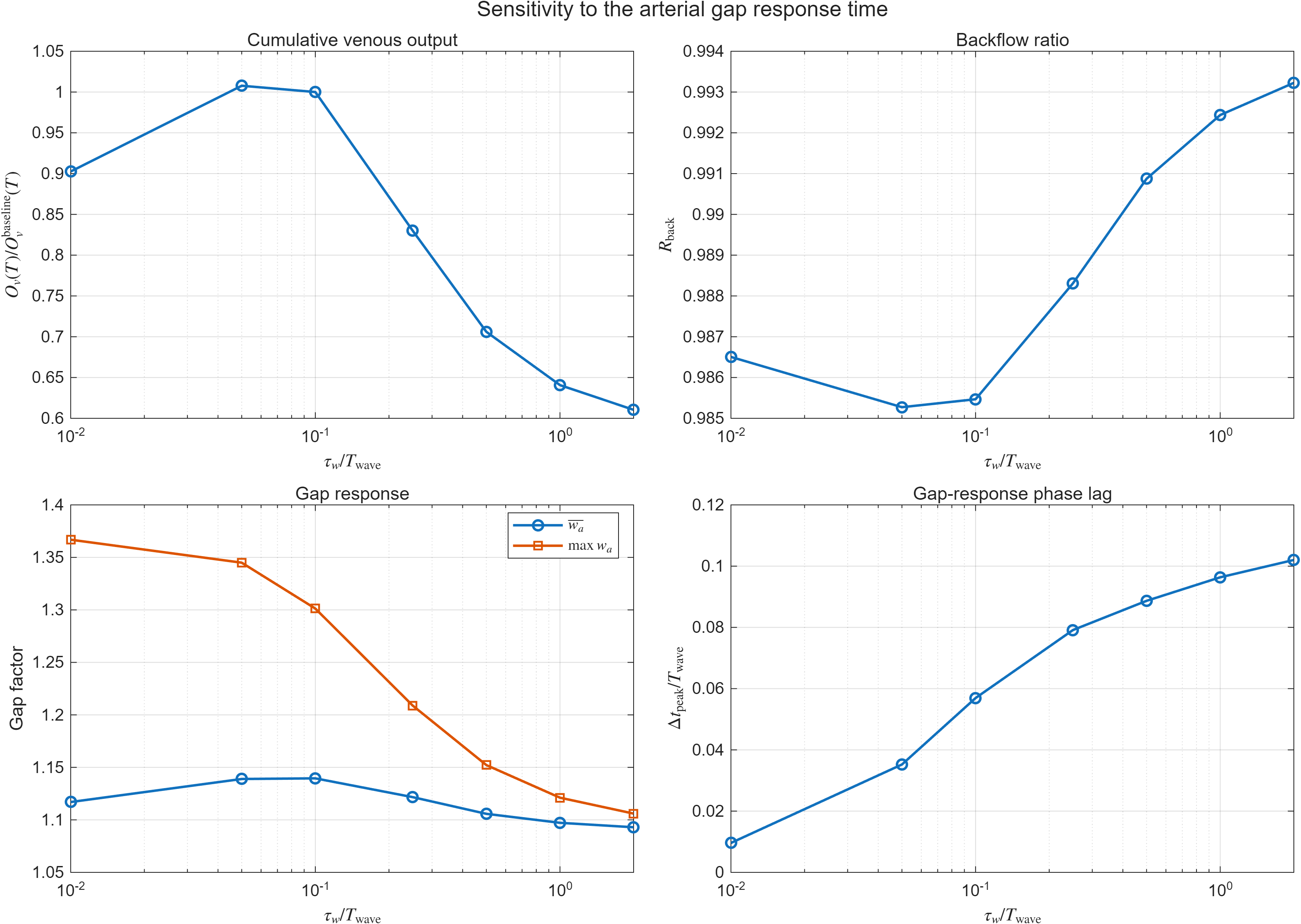}
    \caption{ Sensitivity to the normalized arterial gap-response time under
symmetric slow-vasomotion forcing. Shown are normalized cumulative
venous output, the net directional fraction
\(\eta_{\mathrm{net}}=1-R_{\mathrm{back}}\), mean and maximum gap
factors, and the phase lag between the mechanical PVS source and gap
opening. Clearance is maximized at an intermediate response time,
\(\tau_w/T_{\mathrm{wave}}\approx0.05\text{-}0.1\).}
    \label{fig:tau_gap_sensitivity}
\end{figure}

The influence of the gap relaxation time is examined in
Fig.~\ref{fig:tau_gap_sensitivity}, with quantitative metrics summarized
in Table~\ref{tab:tau_gap_sensitivity}. The response time is normalized
by the vascular-waveform period as
\(\tau_w/T_{\mathrm{wave}}\).

Cumulative venous output depends nonmonotonically on the normalized
response time. The largest output occurs at
\(\tau_w/T_{\mathrm{wave}}=0.05\), while the baseline value
\(\tau_w/T_{\mathrm{wave}}=0.1\) produces an output only about \(1\%\)
below this maximum. The baseline choice therefore lies within a broad
high-clearance regime. It is not a narrowly tuned
parameter value.

A nearly instantaneous response does not maximize clearance. At
\(\tau_w/T_{\mathrm{wave}}=0.01\), the maximum gap factor is the largest
among the tested cases, but both the mean gap factor and cumulative
venous output are lower than at intermediate response times. Thus, peak
gap opening alone does not determine downstream clearance; the duration
and phase of elevated conductance relative to the vascular forcing are
also important.

The range
\(
\tau_w/T_{\mathrm{wave}}\approx0.05\text{-}0.1
\)
combines a relatively large mean gap factor with the smallest
backward-to-forward exchange ratio and the largest net arterial
PVS-to-ECS water transport. Gap modulation
is progressively attenuated at larger response times. Its peak increasingly   lags behind the
mechanical PVS source. Consequently, cumulative venous output decreases
from \(595.20\) at the baseline value to \(314.74\) at
\(\tau_w/T_{\mathrm{wave}}=2\), corresponding to a reduction of
approximately \(47\%\).

\begin{figure}
 \centering
    \subfigure[]{\includegraphics[width=0.45\linewidth]{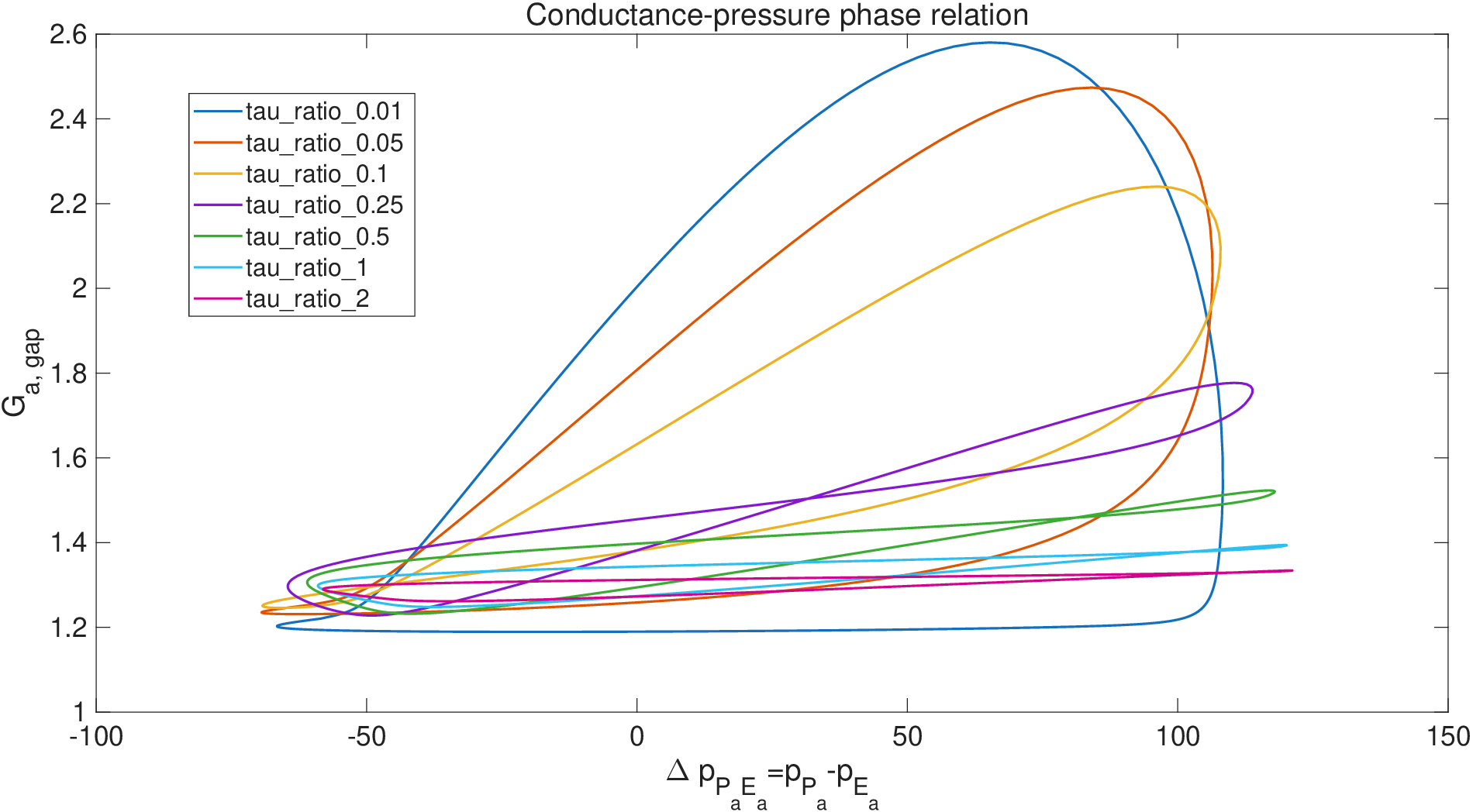}}
    \subfigure[]{\includegraphics[width=0.45\linewidth]{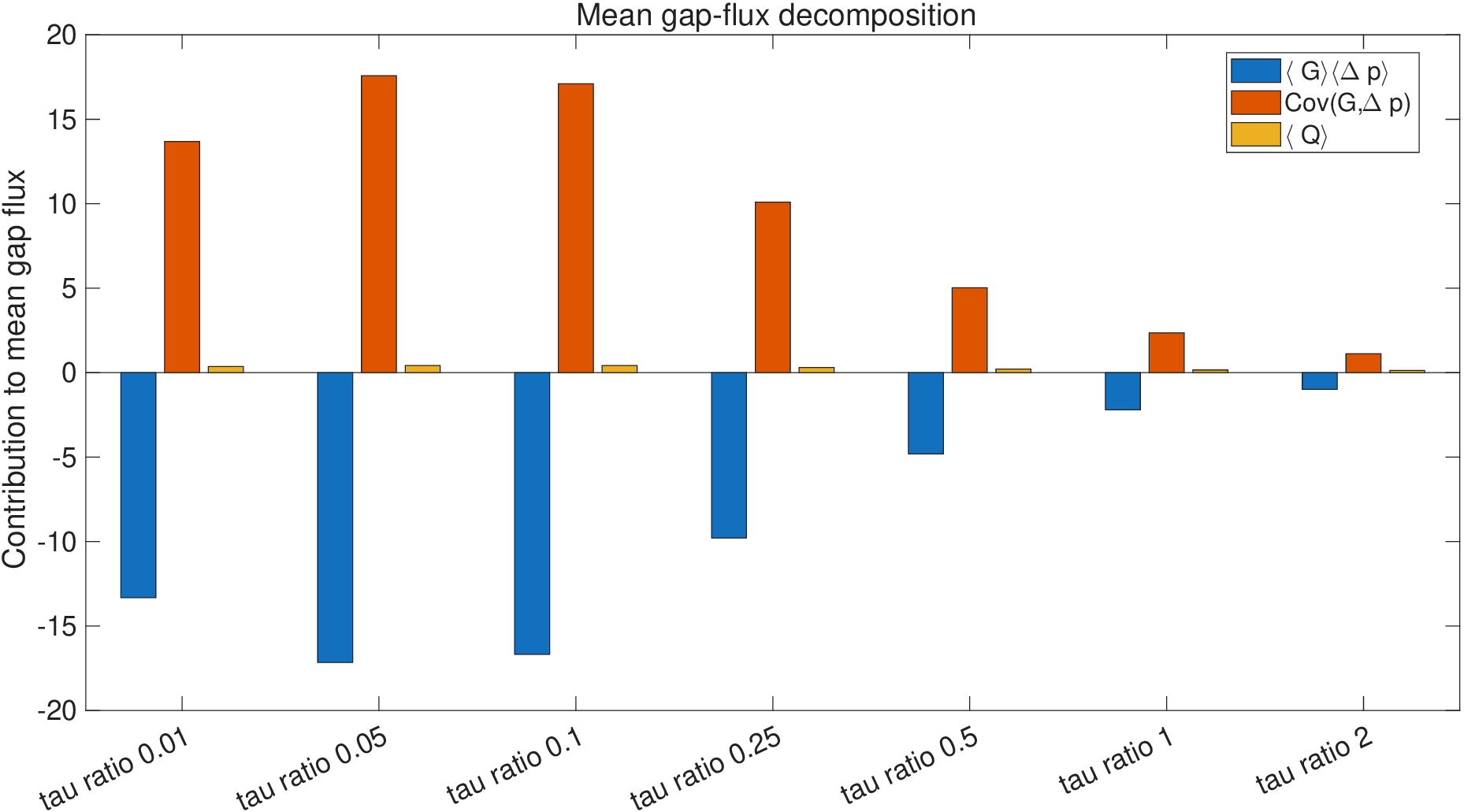}}
\caption{
(a) Conductance-pressure phase relation for the four ablation cases.
(b) Decomposition of the cycle-averaged gap flux.
Mechanical gap regulation generates a positive
conductance-pressure covariance that overcomes the adverse mean-pressure
contribution and produces positive net transport.}
\label{fig:pressurediff-gap-ralation-tau}
\end{figure}

 The peak lag between the gap response and the vascular forcing increases
from approximately \(0.009T_{\mathrm{wave}}\) at
\(\tau_w/T_{\mathrm{wave}}=0.01\) to approximately
\(0.10T_{\mathrm{wave}}\) at
\(\tau_w/T_{\mathrm{wave}}=2\). In Fig. \ref{fig:pressurediff-gap-ralation-tau},  the conductance-pressure phase loops
show how this delay changes the timing of gap opening. At intermediate
response times, large conductance remains preferentially associated
with the positive-pressure, forward-flow portion of the cycle. For slow
gap dynamics, the conductance response is attenuated and extends farther
into the declining and reverse-pressure phases, weakening the distinction
between forward and backward transport.

This mechanism is quantified by decomposing the cycle-averaged gap flux shown in Fig. \ref{fig:pressurediff-gap-ralation-tau}. 
The positive covariance contribution is largest near
\(\tau_w/T_{\mathrm{wave}}=0.05\), where it reaches \(17.58\), and is
similar in the reference case, where it is \(17.10\). It subsequently
decreases to \(1.12\) at
\(\tau_w/T_{\mathrm{wave}}=2\). The resulting cycle-averaged flux
decreases from \(0.424\) near the optimum to \(0.134\) at the slowest
response time.

The intermediate optimum therefore cannot be explained by gap magnitude
alone. Instead, it reflects favorable temporal alignment between gap
conductance and the forward-pressure phase. A small but finite response
delay strengthens hydraulic rectification, whereas excessive delay and
attenuation reduce the conductance-pressure covariance and diminish
net clearance.

\begin{table}[htbp]
\centering
\small
\caption{Sensitivity to the normalized arterial gap-response time under
symmetric slow-vasomotion forcing. Here,
\(O_v^0=O_v(T)\) at
\(\tau_w/T_{\mathrm{wave}}=0.1\), and
\(\eta_{\mathrm{net}}=1-R_{\mathrm{back}}\).}
\label{tab:tau_gap_sensitivity}
\begin{tabular}{cccccccc}
\hline
\(\tau_w/T_{\mathrm{wave}}\)
& \(O_v(T)\)
& \(O_v/O_v^0\)
& \(J_{P_aE_a}^{\mathrm{net}}\)
& \(\eta_{\mathrm{net}}\) (\%)
& \(\overline{w_a}\)
& \(\max w_a\)
& \(\Delta t_{\mathrm{peak}}/T_{\mathrm{wave}}\)\\
\hline
0.01 & 525.47 & 0.883 & 3037.7 & 1.270 & 1.1218 & 1.3716 & 0.0086\\
0.05 & 601.42 & 1.010 & 3451.0 & 1.351 & 1.1465 & 1.3525 & 0.0344\\
0.10 & 595.20 & 1.000 & 3416.6 & 1.336 & 1.1466 & 1.3085 & 0.0569\\
0.25 & 470.56 & 0.791 & 2669.1 & 1.131 & 1.1239 & 1.2112 & 0.0805\\
0.50 & 381.70 & 0.641 & 2066.4 & 0.940 & 1.1040 & 1.1506 & 0.0928\\
1.00 & 335.79 & 0.564 & 1729.8 & 0.821 & 1.0932 & 1.1172 & 0.0964\\
2.00 & 314.74 & 0.529 & 1567.1 & 0.760 & 1.0879 & 1.1009 & 0.0976\\
\hline
\end{tabular}
\end{table}

\subsection{AQP4 redistribution at fixed total abundance}
\label{sec:AQP4_redistribution}
AQP4 is strongly enriched in the perivascular astrocytic endfoot membranes
in the healthy brain. Quantitative immunogold measurements have shown
that AQP4 density in vessel-facing endfoot membranes is approximately
five- to ten-fold higher than that in parenchymal astrocytic membranes
in mice, although the degree of enrichment is lower in humans
\cite{Eidsvaag2017}.

Loss of perivascular AQP4 enrichment does not necessarily imply a
corresponding reduction in total cellular AQP4 expression. In
$\alpha$-syntrophin-deficient mice, total cerebral AQP4 expression was
largely preserved, whereas AQP4 was markedly reduced at perivascular
endfeet and increased in membranes facing the neuropil
\cite{Neely2001}. Similarly, in the Tg-ArcSwe mouse model of
Alzheimer's disease, AQP4 was lost from perivascular endfoot membranes
but increased in fine astrocytic processes surrounding amyloid plaques
\cite{Yang2011}. These observations motivate an idealized
redistribution calculation in which the total functional
AQP4-associated conductance is fixed  while its allocation between
the PVS-facing and ECS-facing endfoot membranes is varied. 

The healthy polarized reference state is represented by
\(\rho_{\mathrm{AQP4}}=0.9\) \cite{Eidsvaag2017}, which gives
\[
L_{P_aA}^{\mathrm{eff}}=0.05,
\qquad
L_{AE_a}^{\mathrm{eff}}=0.01.
\]
Thus, the PVS-facing AQP4-associated conductance is nine times the
ECS-facing contribution in the healthy reference state, while the
background conductances on the two membrane domains are identical.

\begin{figure}
    \centering
    \includegraphics[width=0.75\linewidth]{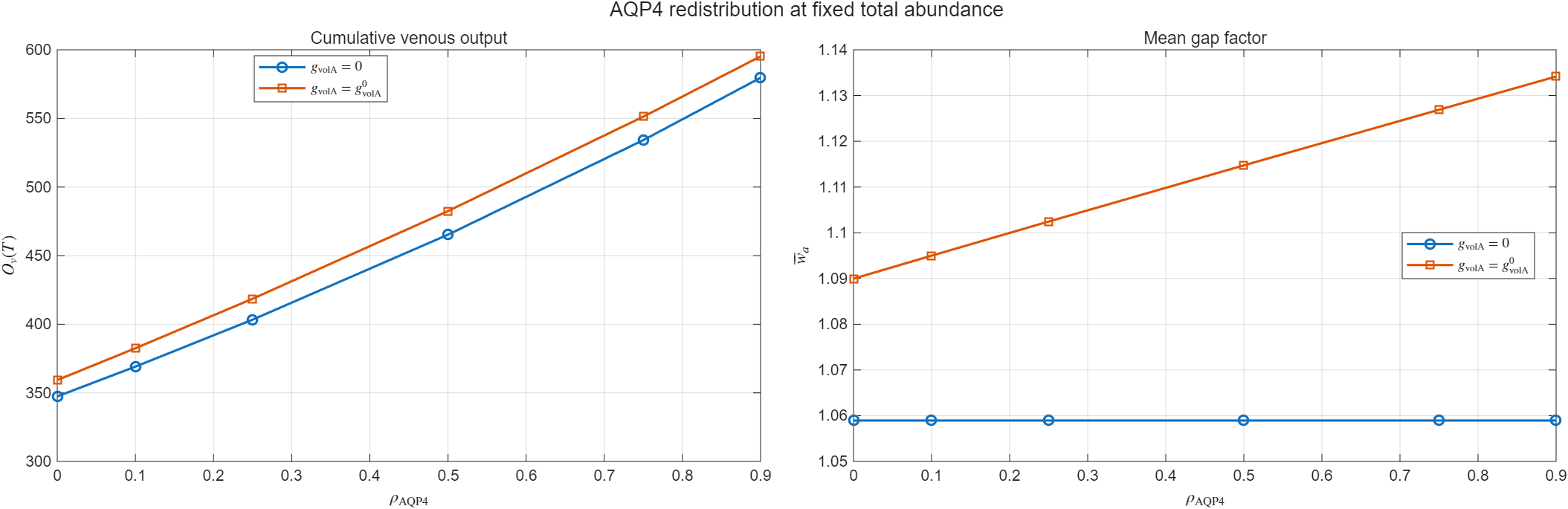}
    \caption{Effect of AQP4 redistribution at fixed total functional abundance.
(a) Cumulative venous output \(O_v(T)\) increases monotonically with
PVS-facing AQP4 localization.
(b) Mean gap factor \(\overline{w}_a\). When
\(g_{\mathrm{volA}}=0\), gap dynamics are independent of AQP4
redistribution. Under the baseline volume-to-gap coupling, increasing
PVS-facing polarization produces an additional increase in the mean
gap factor.}
    \label{fig:AQP4_redistribution}
\end{figure}

Figure~\ref{fig:AQP4_redistribution} shows that cumulative venous output
increases monotonically with the fraction of AQP4 retained on the
PVS-facing membrane. Under the baseline endfoot-volume-to-gap coupling,
the healthy polarized state,
\(\rho_{\mathrm{AQP4}}=0.9\), produces
\(
O_v(T)=595.20.
\)
Complete redistribution to the ECS-facing membrane,
\(\rho_{\mathrm{AQP4}}=0\), reduces the venous output to
\(
O_v(T)=359.32,
\)
corresponding to a \(39.6\%\) reduction. Therefore, the increase in
ECS-facing membrane conductance does not compensate for the loss of
direct PVS-facing hydraulic coupling, even though the total
AQP4-associated conductance is unchanged.

The same qualitative dependence persists when the
endfoot-volume-to-gap feedback is removed. For
\(g_{\mathrm{volA}}=0\), the mean gap factor remains independent of
\(\rho_{\mathrm{AQP4}}\), whereas \(O_v(T)\) decreases from \(579.62\)
at \(\rho_{\mathrm{AQP4}}=0.9\) to \(347.40\) at
\(\rho_{\mathrm{AQP4}}=0\), corresponding to a \(40.1\%\) reduction.
This result demonstrates that the effect of AQP4 redistribution is not
solely caused by dynamic gap-size regulation. Instead, the spatial distribution of AQP4 is important. The AQP4 that faces the PVS affects clearance by modifying the hydraulic pressure and
volume balance of the coupled \(P_a\)-\(A\)-\(E_a\) network.

When the baseline volume-to-gap feedback is retained, the mean gap
factor increases from \(1.08996\) at
\(\rho_{\mathrm{AQP4}}=0\) to \(1.13416\) at
\(\rho_{\mathrm{AQP4}}=0.9\). Thus, AQP4 distribution also affects
clearance indirectly through endfoot-volume-mediated gap regulation.
In the present parameter regime, however, this contribution is smaller
than the direct pressure-network effect.

The qualitative conclusion is independent of  the assumed healthy
ECS-facing conductance. When
\(L_{AE_a}^{\mathrm H}\) is increased from \(0.01\) to \(0.05\),
complete redistribution of the conserved AQP4 pool still reduces
cumulative venous output by \(31.3\%\), compared with \(39.6\%\) under
the reference parameterization. These results suggest that the spatial
localization of AQP4, rather than total AQP4 abundance alone, is an
important determinant of waveform-driven downstream clearance.

The conductance-pressure phase loops show that increasing
\(\rho_{\mathrm{AQP4}}\) systematically shifts the loops upward and
slightly enlarges them, indicating that larger PVS-facing AQP4 abundance
is associated with larger gap conductance throughout the waveform cycle,
especially during the positive-pressure phase. Thus, when forward driving
is favorable, the gap tends to be more open for larger
\(\rho_{\mathrm{AQP4}}\), whereas during the reverse-pressure phase the
conductance remains comparatively smaller. This phase-dependent alignment
enhances hydraulic rectification.

This interpretation is confirmed by the decomposition of the
cycle-averaged gap flux.
As \(\rho_{\mathrm{AQP4}}\) increases, the adverse mean-pressure term
\(\langle G\rangle\langle \Delta p\rangle\) becomes slightly more
negative, but this is more than compensated by the increase in the
positive covariance term. Consequently, the net cycle-averaged
gap-mediated flux increases with \(\rho_{\mathrm{AQP4}}\), consistent
with the observed increase in cumulative venous output.

These results support the interpretation that PVS-facing AQP4 improves
clearance primarily by strengthening the hydraulic coupling, rather than by directly
carrying the dominant water flux.

\begin{figure}[!htb]
 \centering
    \subfigure[]{\includegraphics[width=0.45\linewidth]{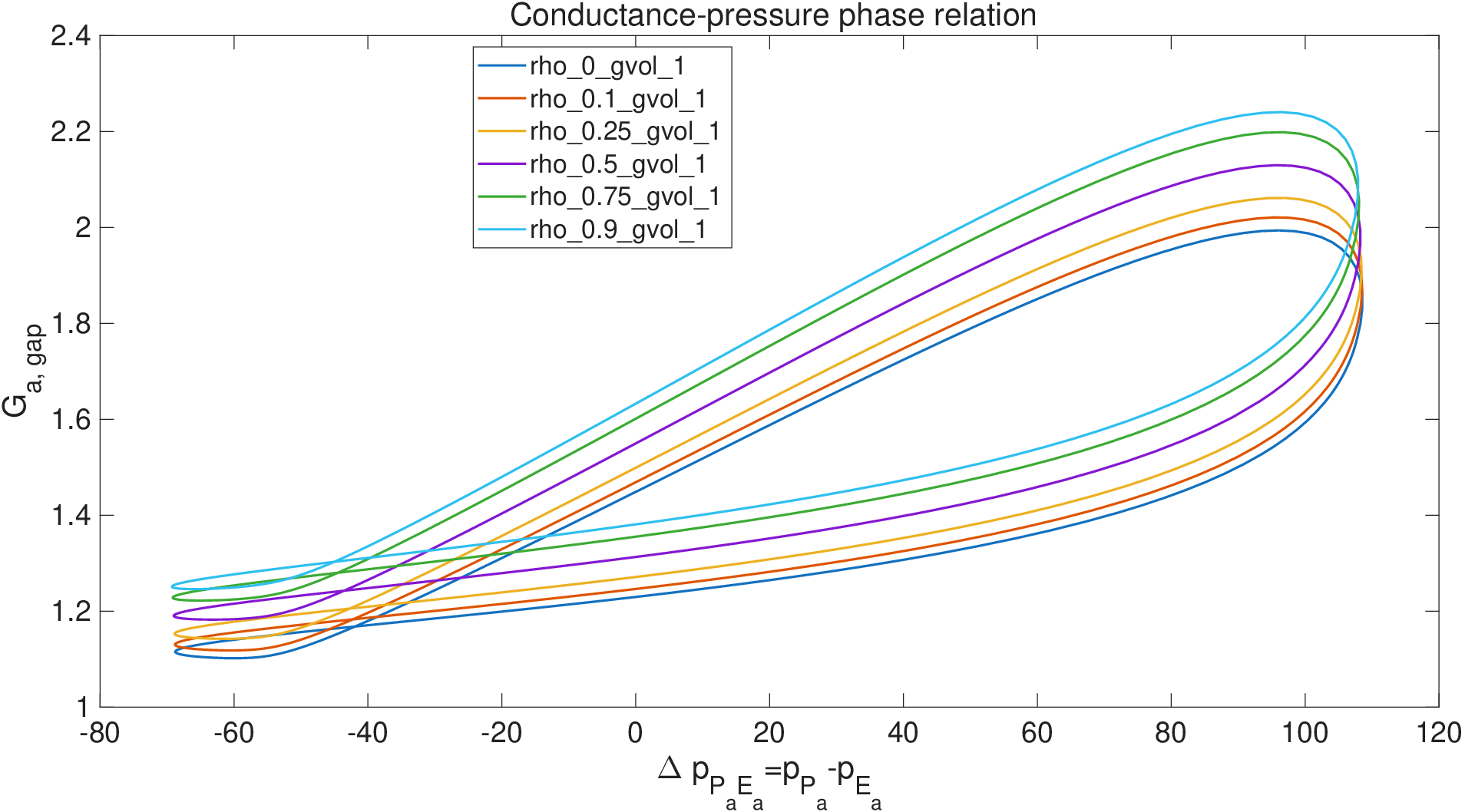}}
    \subfigure[]{\includegraphics[width=0.45\linewidth]{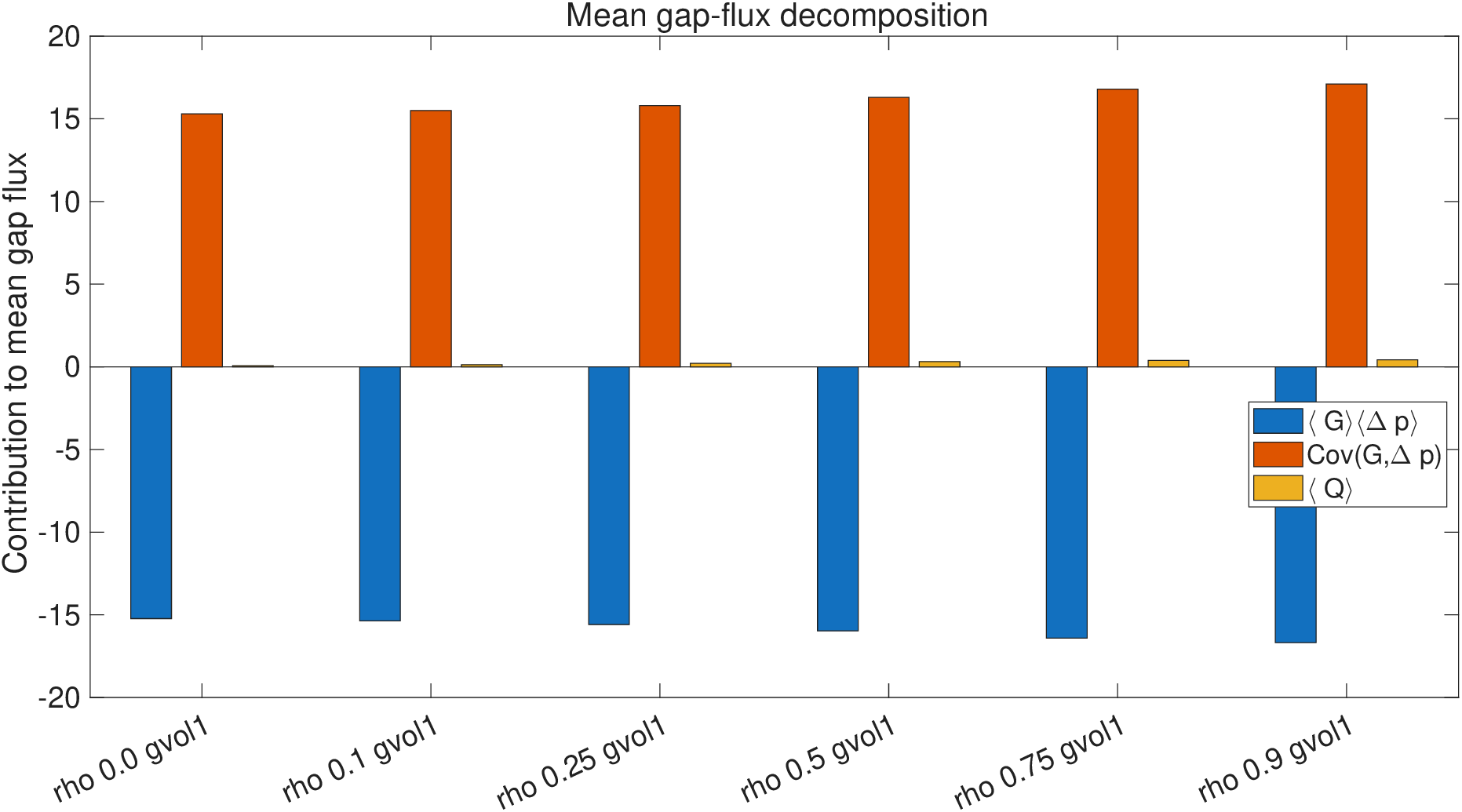}}
\caption{Mechanistic interpretation of AQP4 redistribution under the full
dynamic-gap model (\(g_{\mathrm{volA}}=g_{\mathrm{volA}}^0\)).
(a) Conductance-pressure phase relation.
(b) Decomposition of the cycle-averaged gap flux.
 }
\label{fig:pressurediff-gap-ralation-rho}
\end{figure}

\subsection{Aging-associated impairment of glymphatic clearance}

We next investigated how aging-associated alterations in vascular motion,
PVS mechanics, and effective perivascular AQP4 function affect
waveform-driven clearance. An idealized symmetric slow-vasomotion waveform
was used as the vascular forcing. 
Three independent aging-related parameters were selected to represent
three distinct components of the gliovascular interface: vascular forcing,
mechanical transmission through the PVS, and astrocytic water regulation.

First, aging is associated with reduced intracortical arterial-wall
deformation and decreased vascular compliance, whereas arterial-wall motion
is an important driver of perivascular CSF transport
\cite{kress2014impairment,mestre2018flow}. We therefore use the normalized
vessel-motion amplitude \(\epsilon/\epsilon_0\) to represent the strength
of vascular mechanical forcing, with \(\epsilon/\epsilon_0<1\)
corresponding to reduced vessel-wall motion.

Second, aging and neurodegenerative disease are accompanied by structural
and compositional remodeling of the periarterial PVS and its surrounding
pial and extracellular-matrix structures \cite{mestre2022periarteriolar}. Recent
high-fidelity mechanical simulations further suggest that changes in PVS
stiffness can strongly reduce, eliminate, or even reverse fluid exchange at
the gliovascular interface \cite{Causemann2026}. In the present reduced
geometry, the parameter \(\eta_o\) controls how strongly the outer PVS
boundary follows vessel-wall motion. Increasing \(\eta_o\) reduces the
effective compression of the PVS during vessel dilation and is therefore
used as a phenomenological representation of aging-associated alteration
in PVS mechanical coupling. Note that \(\eta_o\) is an effective
model parameter rather than a directly measured age-dependent material
property.

Third, aging is associated with loss of perivascular AQP4 polarization,
and both the presence of AQP4 and its localization at astrocytic endfeet
contribute to extracellular solute clearance
\cite{kress2014impairment,bojarskaite2024role,khan2026impaired}. We therefore introduce the
effective perivascular AQP4 factor \(\rho_{\rm AQP4}\), with
\(\rho_{\rm AQP4}<1\) representing reduced AQP4 availability,
polarization, or functional water permeability.

\begin{figure}[htp]
    \centering
  \subfigure[] {\includegraphics [width=0.65\linewidth]{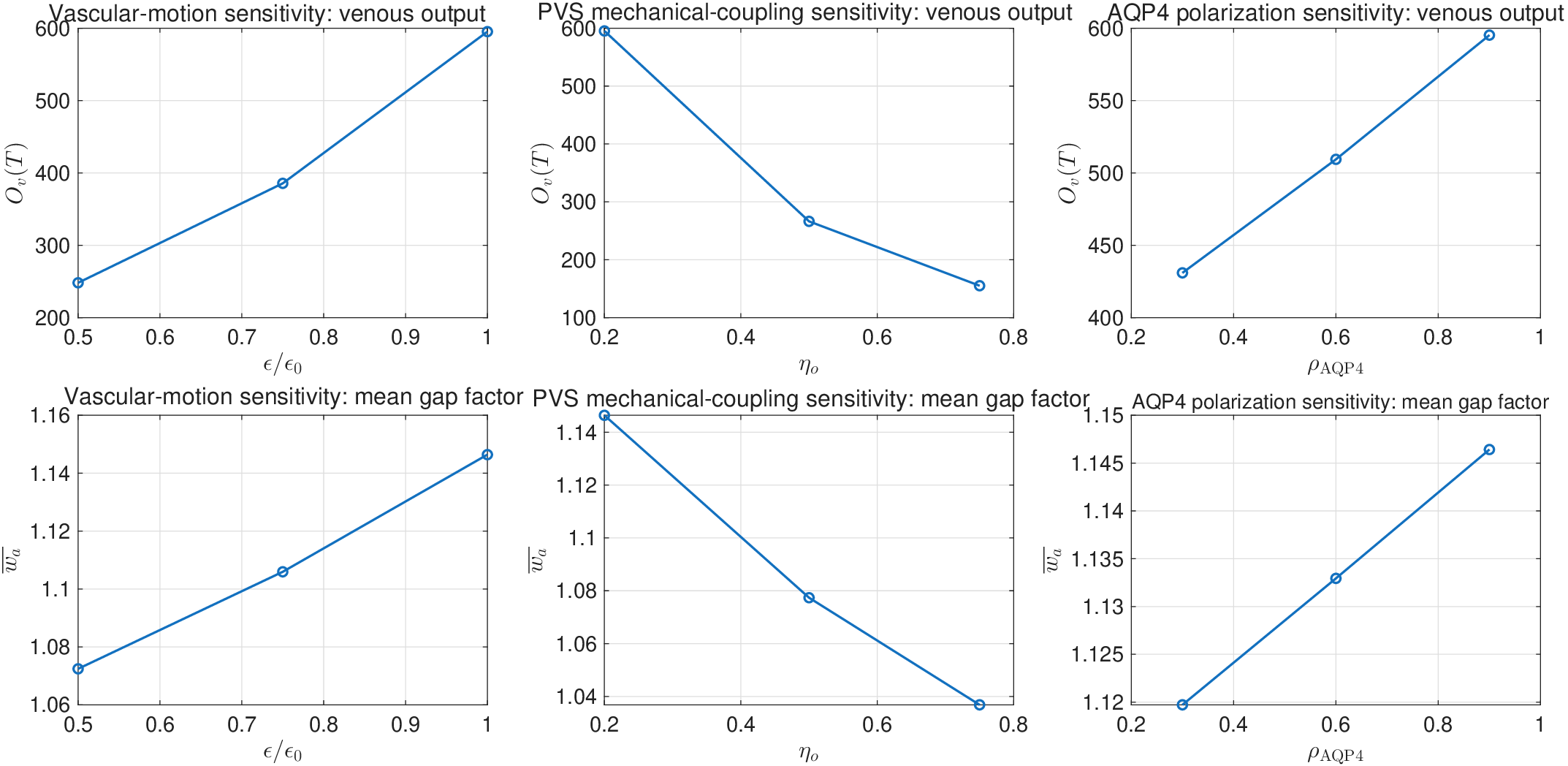}}
    \subfigure[] {\includegraphics [width=0.65\linewidth]{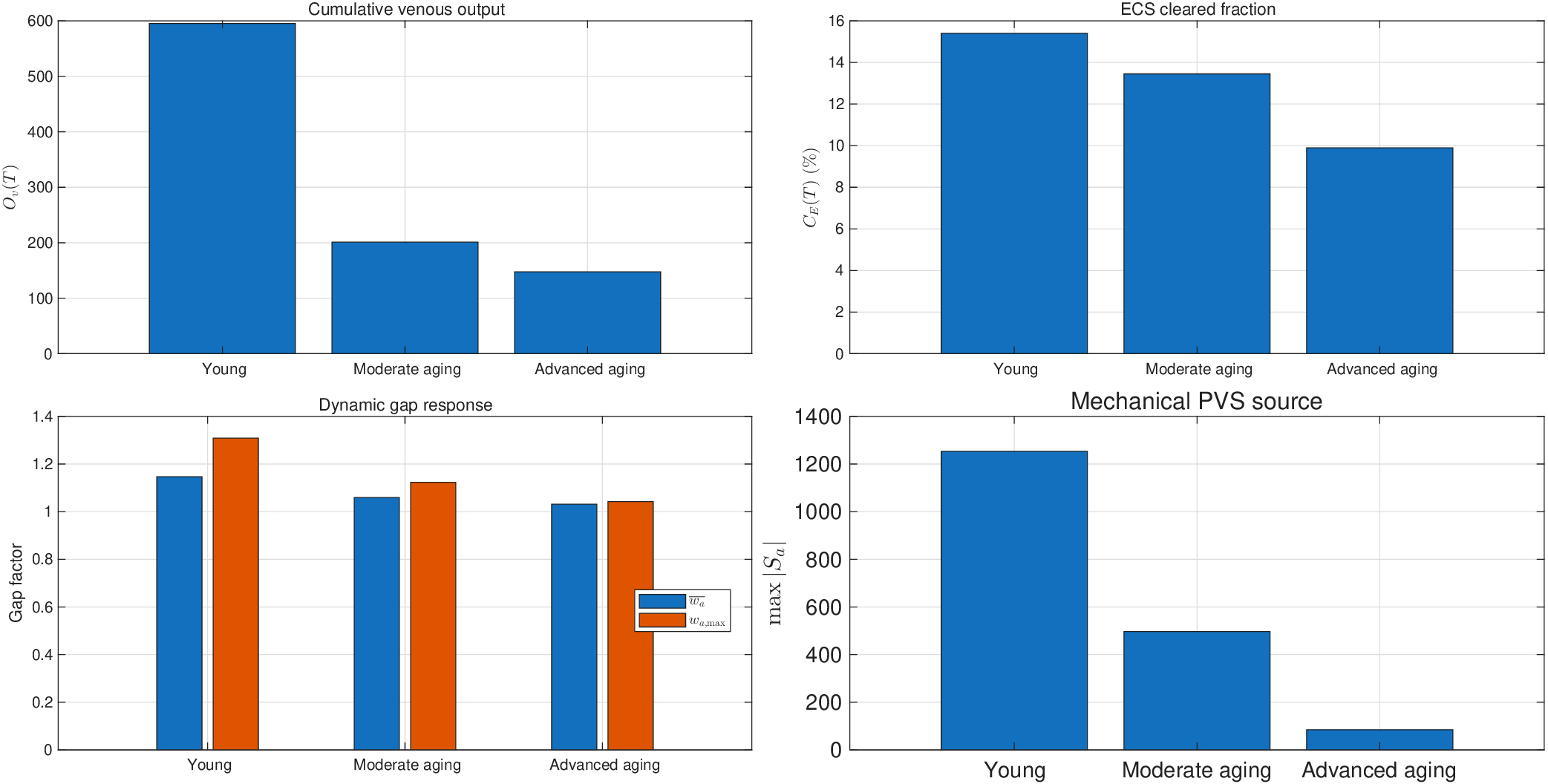}}
        \subfigure[] {\includegraphics [width=0.65\linewidth]{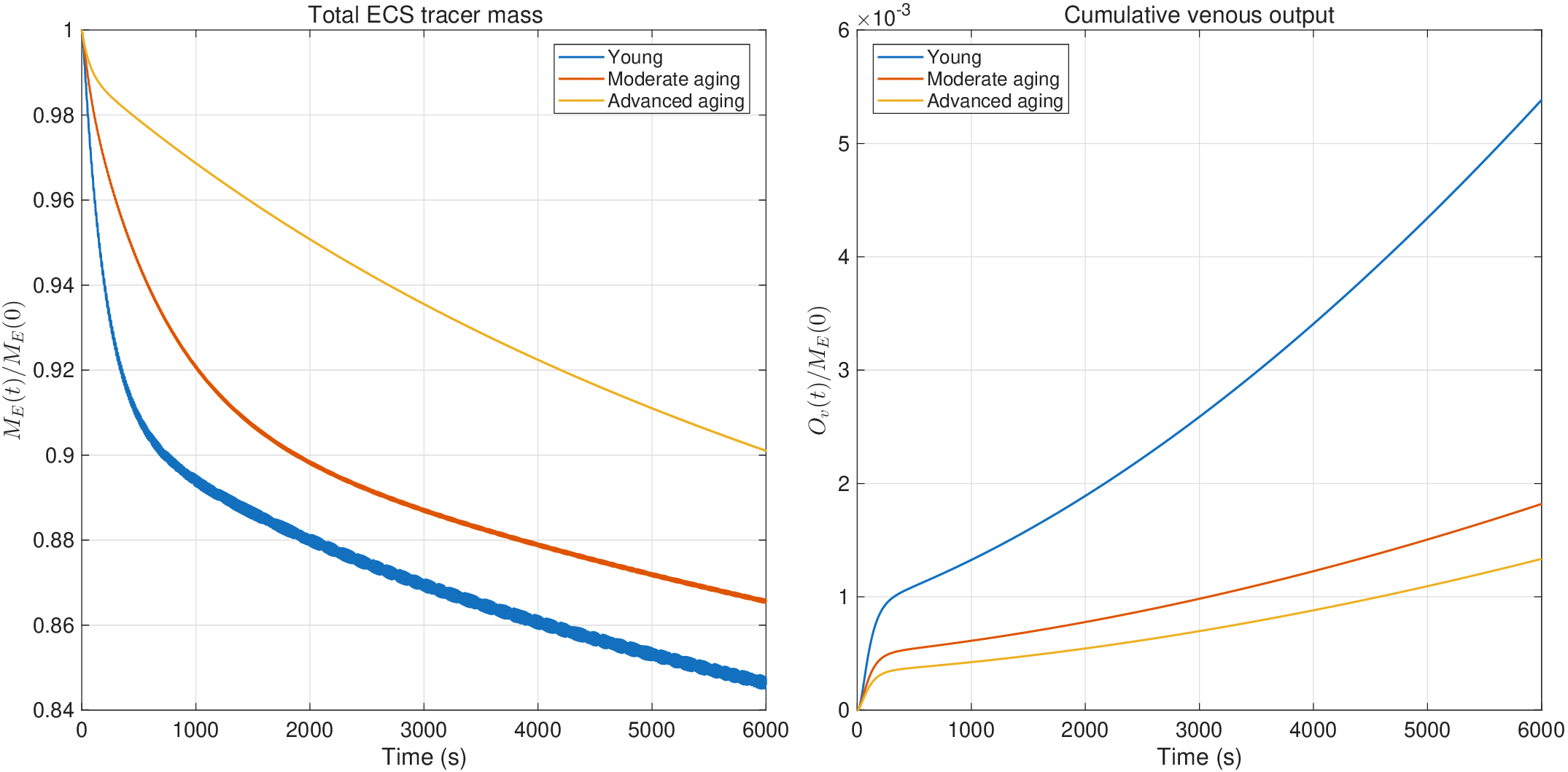}}
    \caption{Aging-associated impairment under symmetric slow-vasomotion forcing.
(a) One-factor sensitivity to vessel-motion amplitude, outer-PVS
boundary coupling, and AQP4 polarization.
(b) Representative young, moderate aging-like, and advanced aging-like
phenotypes.
(c) Normalized ECS tracer mass and cumulative venous output.
Total functional AQP4 conductance is fixed, while aging-related
depolarization is represented by decreasing
\(\rho_{\rm AQP4}\).}
    \label{fig:aging}
\end{figure}

 The combined aging-like phenotypes produce a substantial compounded
impairment, as shown in Fig.~\ref{fig:aging}(b) and
Table~\ref{tab:aging_summary}. In these simulations, the total
functional AQP4-associated conductance is held fixed, while aging-related
loss of perivascular polarization is represented by decreasing
\(\rho_{\mathrm{AQP4}}\). Relative to the young-reference case, the
moderate aging-like phenotype reduces the maximum mechanical PVS source
by approximately \(60\%\), whereas the advanced aging-like phenotype
reduces it by approximately \(93\%\). The corresponding excess mean gap
opening, measured by \(\overline{w_a}-1\), decreases by approximately
\(60\%\) and \(79\%\), respectively.

These changes strongly suppress gap-mediated transport. Net arterial
PVS-to-ECS water exchange decreases by approximately \(68\%\) in the
moderate aging-like case and by approximately \(89\%\) in the advanced
aging-like case. Cumulative venous output decreases from \(595.2\) in
the young-reference case to \(201.2\) and \(147.6\), corresponding to
reductions of approximately \(66\%\) and \(75\%\), respectively. Venous
outlet efficiency also decreases from \(9.03\%\) to \(4.17\%\) and
\(3.32\%\). Thus, aging-associated impairment affects both transport
toward the venous PVS and subsequent downstream removal.

These results support a  reduced composite mechanism in which weakened
vascular motion, altered transmission of vessel deformation to the PVS,
and loss of perivascular AQP4 polarization act through distinct but
compounded pathways. Reduced vascular and PVS deformation weakens the
mechanical forcing and dynamic gap response, whereas AQP4
depolarization redistributes a fixed total AQP4-associated conductance
from the PVS-facing to the ECS-facing membrane and thereby alters the
hydraulic balance of the gliovascular interface.

The total ECS tracer mass in Fig.~\ref{fig:aging}(c) exhibits a rapid
initial decrease followed by a slower long-time decay. However, the ECS
depleted fraction remains much larger than the normalized cumulative
venous output. For example, the final ECS depleted fractions are
approximately \(15.40\%\), \(13.45\%\), and \(9.90\%\) in the young,
moderate aging-like, and advanced aging-like cases, whereas the
corresponding venous output fractions are only \(0.538\%\), \(0.182\%\),
and \(0.133\%\). Early ECS tracer loss therefore includes redistribution
into the arterial and venous PVS compartments and should not be
interpreted as complete downstream clearance. This distinction motivates
the simultaneous reporting of ECS depletion, cumulative venous output,
and venous outlet efficiency.

\begin{table}[htbp]
\centering
\caption{Clearance and gap-regulation metrics for representative
aging-like phenotypes under symmetric slow-vasomotion forcing. Total
functional AQP4-associated conductance is fixed at
\(\alpha_{\rm AQP4}=1\), while loss of perivascular polarization is
represented by decreasing \(\rho_{\rm AQP4}\).}
\label{tab:aging_summary}
\begin{tabular}{lcccccccc}
\hline
Case
& $\epsilon/\epsilon_0$
& $\eta_o$
& $\rho_{\rm AQP4}$
& $\overline{w_a}$
& $C_E(T)$
& $F_v(T)$
& $\eta_v(T)$
& $F_{P_aE_a}^{\rm net}$ \\
\hline
Young
& 1.00 & 0.20 & 0.90
& 1.146
& $15.40\%$
& $0.538\%$
& $9.027\%$
& 3416.1 \\

Moderate aging-like
& 0.75 & 0.50 & 0.60
& 1.059
& $13.45\%$
& $0.182\%$
& $4.173\%$
& 1092.2 \\

Advanced aging-like
& 0.50 & 0.75 & 0.30
& 1.031
& $9.90\%$
& $0.133\%$
& $3.318\%$
& 386.8 \\
\hline
\end{tabular}
\end{table}

 \section{Discussion}\label{sec:discussion}

The present study develops a reduced multicompartment framework for examining how vascular forcing, PVS deformation, inter-endfoot gap regulation, and AQP4-mediated water exchange jointly regulate perivascular-interstitial transport. The model is not intended to reproduce the complete brain-wide glymphatic system. Rather, it provides a mechanistic description of how local processes at the gliovascular interface can generate directional transport and influence downstream tracer clearance.

The results support several main conclusions. Large oscillatory water exchange does not necessarily imply effective net clearance. Temporal asymmetry in vascular forcing can rectify otherwise bidirectional exchange by suppressing recovery-phase backflow. Dynamic regulation of the inter-endfoot gap provides an additional rectification mechanism whose effectiveness depends strongly on the timing of gap opening relative to the PVS-ECS pressure difference. Finally, although direct AQP4-mediated hydrostatic flux is much smaller than gap-mediated flux, PVS-facing AQP4 can substantially influence clearance by modifying the pressure-volume balance that drives the dominant gap pathway.

\paragraph{Waveform-dependent rectification and dynamic gap regulation.}
The distinction between large local exchange and net directional transport is central to the present results. Cardiac-like forcing generates substantial forward and backward motion, but the two components largely cancel over a cycle. In contrast, slow vasodilation, particularly when temporally asymmetric, produces stronger net PVS-to-ECS transport because the recovery phase generates less reverse flow. Thus, vascular signals with similar amplitudes can produce very different transport outcomes depending on their temporal structure.

Dynamic gap regulation provides a second mechanism for rectification. Because the hydraulic conductance of a narrow inter-endfoot cleft is highly sensitive to its effective width, modest gap changes can strongly modulate PVS-ECS exchange. The key factor, however, is not simply the magnitude of gap opening. Net transport is enhanced when increased conductance is preferentially aligned with the forward-pressure phase and reduced during the reverse-pressure phase. In terms of the cycle-averaged flux,
  a positive contribution from the time-dependent conductance-pressure alignment can overcome an unfavorable mean pressure difference and generate net forward transport. The response-time analysis further shows that this alignment is strongest when the gap responds rapidly, but not instantaneously, to vascular forcing. This suggests that hydraulic rectification depends on the magnitude, duration, and timing of gap opening rather than on gap width alone.

Waveform asymmetry and dynamic gap regulation therefore act as partially substitutable rectification mechanisms. Dynamic gap modulation provides a large relative enhancement when the vascular waveform itself is nearly symmetric, whereas its additional effect is smaller when temporal asymmetry has already strongly suppressed recovery-phase backflow.

\paragraph{AQP4 regulation in a gap-dominated transport system.}
The model provides a possible resolution of the apparent tension between experiments showing impaired glymphatic transport after AQP4 disruption and mechanical simulations suggesting that pressure-driven water exchange occurs predominantly through inter-endfoot gaps. In the present model, the gap-mediated flux is substantially larger than the direct PVS-facing endfoot flux, so hydrostatic exchange remains gap dominated. Nevertheless, AQP4 alters the hydraulic state of the coupled $P_a$-$A$-$E_a$ system and thereby modifies the pressure difference driving the gap pathway.

This distinction is particularly clear in the AQP4 redistribution experiment. When total functional AQP4-associated conductance is held fixed, moving AQP4 from the PVS-facing membrane toward the ECS-facing membrane still reduces downstream venous output. The compensatory increase in ECS-facing conductance therefore does not reproduce the hydraulic effect of PVS-facing AQP4. Endfoot-volume-dependent gap regulation provides an additional contribution, but the redistribution results remain qualitatively unchanged when this feedback is removed. These findings suggest that perivascular AQP4 localization can be important independently of total AQP4 abundance.

The present framework is complementary to geometry-resolved models of the gliovascular interface. High-fidelity simulations resolve local deformation, pressure fields, and flow through individual inter-endfoot clefts, whereas the reduced model sacrifices spatial detail in order to follow transport over longer times and through an arterial-ECS-venous network. This reduction makes it possible to distinguish local bidirectional exchange from net transport, venous PVS storage, and downstream clearance, and to examine how AQP4 localization and time-dependent gap conductance influence these quantities.

\paragraph{Compounded impairment with aging.}
The aging-like simulations support a multi-hit interpretation of impaired glymphatic transport. Reduced vascular-wall motion weakens the mechanical driving force, altered PVS mechanical coupling reduces the transmission of vascular deformation to the surrounding fluid space, and loss of perivascular AQP4 polarization modifies the hydraulic balance at the gliovascular interface. Because these perturbations act through different components of the system, their combined effect produces a substantial reduction in gap-mediated transport and downstream clearance.

These simulations should be interpreted mechanistically rather than as direct representations of chronological aging. Parameters such as the outer-boundary coupling factor and effective AQP4 polarization are reduced-model quantities rather than directly measured age-dependent material properties. Their role is to isolate how several experimentally observed aging-associated changes may combine within a common transport network.

\paragraph{Tracer redistribution and downstream clearance.}
The model also emphasizes that loss of tracer from the ECS is not equivalent to removal from the system. Tracer may leave the ECS and accumulate transiently in the arterial or venous PVS before reaching the downstream outlet. Measures of ECS depletion, venous PVS storage, cumulative venous output, and outlet efficiency therefore characterize different stages of transport. This distinction may be important when comparing experiments that quantify changes in tissue fluorescence with measurements of downstream drainage.

\paragraph{Limitations and future directions.}
The model is spatially lumped and treats each PVS, ECS, and endfoot region as a well-mixed compartment. It therefore cannot resolve axial pressure gradients, local recirculation, heterogeneous gap distributions, vascular branching, or three-dimensional tissue deformation. The prescribed vascular waveforms and the dynamic-gap constitutive law are also idealized, and the gap-response parameters should be regarded as effective quantities rather than directly measured microscopic properties.

Dynamic ionic and osmotic effects are not included. The present study isolates hydrostatic and mechanical coupling and does not imply that osmotic forces, ion transport, or astrocyte volume regulation are physiologically negligible. These mechanisms may become particularly important during neural activation, edema, ischemia, and other pathological conditions.

The AQP4 redistribution model is also simplified, since displaced AQP4 is represented only through redistribution between PVS-facing and ECS-facing membrane domains. In vivo, loss of perivascular polarization may redistribute AQP4 to fine astrocytic processes and the soma. A natural extension would therefore couple the endfoot compartment to a dynamic intracellular soma/process reservoir and include non-perivascular AQP4-mediated water exchange.

Future work should combine the present reduced model with experimentally measured vascular waveforms and geometry-resolved simulations to estimate effective conductances, pressure-volume relations, and gap-response parameters. Simultaneous measurements of vascular motion, PVS geometry, endfoot deformation, AQP4 localization, gap dynamics, and downstream tracer removal would provide particularly valuable tests of the proposed rectification mechanisms.

\section{Conclusions}\label{sec:conclusion}


The study provides a possible resolution of the apparent discrepancy
between experiments showing substantial clearance impairment after AQP4
disruption and mechanical models predicting a small direct
pulsation-driven transmembrane AQP4 flux.

PVS-facing AQP4 transport need
not carry the dominant PVS-ECS water flux in order to influence
clearance. It can alter the pressure-volume balance of the
\(P_a\)-\(A\)-\(E_a\) network and thereby modify the hydraulic driving
force of the dominant gap pathway. Endfoot-volume changes provide a
second, more modest mechanism by modulating gap conductance. These
mechanisms are distinct but interacting, and their relative importance
depends on the vascular waveform and the timescale of gap regulation.

The mechanism decomposition analysis identifies vascularly induced
mechanical opening as the primary driver of dynamic-gap enhancement,
whereas endfoot-volume feedback provides secondary regulation. The
response-time analysis further predicts a high-clearance regime when
the effective gap-response time is approximately \(5\%\)-\(10\%\) of
the vascular-waveform period. Thus, transport is controlled not only by
how far the gap opens, but also by when and for how long the elevated
conductance is maintained.

Finally, the aging-associated simulations support a multi-hit
interpretation of glymphatic impairment. Reduced vessel-wall motion,
altered transmission of deformation through the PVS, and impaired
effective PVS-facing AQP4 transport act through different components of
the system to suppress mechanical forcing, gap-mediated exchange, and
venous-directed clearance.

The dynamic-gap mechanism proposed in this study is a
physiologically motivated, model-based hypothesis rather than an
experimentally validated microscopic law. Its value lies in showing
that the available experimental and computational observations can be
reconciled within a mechanically consistent framework and in generating
specific, falsifiable predictions. In particular, the model is meant to motivate new experiments
  that simultaneously measure vascular motion, PVS geometry,
endfoot deformation, inter-endfoot gap dynamics, AQP4 polarization, and
downstream tracer removal. Such experiments will be essential for
determining whether the proposed mechanisms operate in vivo and for
refining the reduced constitutive relations introduced here.

\section*{Acknowledgments}

The authors thank the Fields Institute for Research in Mathematical
Sciences for hosting the research visit through the MADS Laboratory and
for providing a stimulating environment for this work. S.X. would also like to thank Marius Causemann and Marie E. Rognes for
insightful discussions on possible mechanisms through which AQP4 may
regulate glymphatic clearance.

\section*{Declarations}
\paragraph{Ethics approval and consent to participate} 
Not applicable: this manuscript presents computer simulations and does not report on or involve any animals, humans, human data, human tissue or plants.  
\paragraph{Consent for publication}  Not applicable: manuscript does not contain data from any individual person.  

\paragraph{Competing interests}  The authors declare that they have no competing interests.

\paragraph{Availability of data and materials}
The dataset(s) supporting the conclusions of this article is(are) available in the \href{https://github.com/xsxztr/pvs-endfeet-ecs-multicompartment-model.git}{https://github.com/xsxztr/pvs-endfeet-ecs-multicompartment-model.git}
\bibliographystyle{plain}
\bibliography{reference}

\end{document}